\documentclass[twocolumn]{aastex631}

\newcommand{\civ}{C {\sc iv}}
\newcommand{\heii}{He {\sc ii}}
\newcommand{\ciii}{C {\sc iii}}

\newcommand{\oiii}{O {\sc iii}}

\newcommand{\mgii}{Mg {\sc ii}}

\shorttitle{AASTeX v6.3.1 Sample article}
\shortauthors{Zhang et al.}
\usepackage{color}
\usepackage{amsmath}
\usepackage{amssymb}
\usepackage{float}
\usepackage{changepage}
\usepackage{multirow}
\usepackage{threeparttable}
\graphicspath{{./}{figures/}}

\begin{document}

\title{Constraining the nature of active galactic nuclei through circumgalactic Ly$\alpha$ emission at $z=2-3$}

\author{Shiwu Zhang}
\affiliation{Department of Automation, Tsinghua University, Beijing 100084, China}
\author{Zheng Cai}
\affiliation{Department of Astronomy, Tsinghua University, Beijing 100084, China}
\author{Fabrizio Arrigoni Battaia}
\affiliation{Max-Planck-Institut f$\ddot{u}$r Astrophysik, Karl-Schwarzschild-Stra$\beta$e 1, Garching bei M$\ddot{u}$nchen, D-85748, Germany}
\author{Sebastiano Cantalupo}
\affiliation{Department of Physics, University of Milan Bicocca, Milano, 20126, Italy}
\author{Mingyu Li}
\affiliation{Department of Astronomy, Tsinghua University, Beijing 100084, China}
\author{Ben Wang}
\affiliation{Department of Astronomy, Tsinghua University, Beijing 100084, China}
\author{Yuduo Guo}
\affiliation{Department of Automation, Tsinghua University, Beijing 100084, China}
\author{Aura Obreja}
\affiliation{Interdisziplin\"ares Zentrum f\"ur Wissenschaftliches Rechnen, Universit\"at Heidelberg, Heidelberg, D-69120, Germany}
\author{Haibin Zhang}
\affiliation{National Astronomical Observatory of Japan, Tokyo, 181-8588, Japan}
\author{Zihao Li}
\affiliation{Niels Bohr Institute, University of Copenhagen, Copenhagen N, DK-2200, Denmark}
\author{Donghui Quan}
\affiliation{Department of Physics, Xi'an Jiaotong-Liverpool University, SuZhou 215123, China}



\begin{abstract}
We present a comprehensive analysis of circumgalactic Ly$\alpha$ nebulae around 59 unobscured and 26 obscured quasars at $z=2-3$, observed with the Keck Cosmic Web Imager (KCWI), to constrain the nature of active galactic nuclei (AGN) at cosmic noon. 
We find that Ly$\alpha$ nebulae around unobscured quasars are significantly less symmetric having a symmetry parameter of $\alpha_{\rm w}=0.2-0.6$ and more spatially extended having a scale length of $r_{\rm h}=10.7\pm0.5$ kpc than those around obscured quasars ($\alpha_{\rm w}=0.6-0.8$; $r_{\rm h}\approx6.6-7.7$ kpc).
Unobscured quasars also exhibit steeply declining velocity dispersion profiles with the slope of $-4.3\pm0.4$ km s$^{-1}$ kpc$^{-1}$, indicative of large-scale outflows, whereas obscured quasars display flat profiles ($-0.2\pm0.7$ and $-0.6\pm0.4$ km s$^{-1}$ kpc$^{-1}$). 
The degree of quasar obscuration appears to be intrinsically linked to nebular asymmetry and extent, a relationship that could be in tension with the standard orientation-based AGN unified model, as it expects unobscured-quasar nebulae to be more symmetric and compact. 
These results naturally fit the evolutionary scenario, where AGN feedback drives a transition from an obscured to an unobscured phase—progressively redistributing gas to larger radii, introducing anisotropy, and driving turbulence. 
Taken together, our findings favor the evolutionary scenario over the purely orientation-based unified model for quasars at cosmic noon. 
\end{abstract}

\keywords{Quasar; Circumgalactic Medium; Ly$\alpha$ nebulae; high-redshift;}


\section{Introduction} \label{sec:intro}
The majority of baryons associated with galaxies reside not within their stellar bodies, but in the diffuse gas that extends throughout their dark matter halos. 
This reservoir, known as the circumgalactic medium (CGM), plays a central role in galaxy evolution by mediating the exchange of mass, energy, and metals between galaxies and their surroundings. 
The CGM is inherently multiphase, spanning temperatures from $10-10^{7}$ K \citep{Li2021,Zhang2023a,Zhang2024}, and is shaped by a complex interplay of large-scale gas inflows \citep{Martin2015,Kere2005,Decataldo2024}, galactic outflows \citep{Fabian2012,Harrison2024,Vayner2024}, and recycling processes \citep{Zhang2023a,Zhang2023b,Angles2017}. 
These mechanisms collectively regulate star formation and the growth of supermassive black holes (SMBHs), making the CGM a key component in understanding galaxy–SMBH co-evolution \citep{Tumlinson2017}.

In recent years, rapid advances in deep imaging and integral field spectroscopy (IFS) have opened a new window onto the CGM by enabling its direct mapping in emission \citep{Cantalupo2014,Arrigoni2019,Cai2019,Lobos2025,Li2024}.
Since hydrogen constitutes the dominant baryonic component of the CGM, these observations have primarily focused on extended Ly$\alpha$ emission—so-called Ly$\alpha$ nebulae—powered by (i) photoionization and recombination due to strong incident ultraviolet radiation \citep{Farina2019}, (ii) gravitational cooling radiation \citep{Daddi2021}, (iii) shocks driven by energetic outflows \citep{Costa2022}, and (iv) resonant scattering \citep{Sanderson2021}.

Such direct imaging, particularly the spatially resolved spectra provided by IFS, has enabled a range of fundamental measurements: imaging large-scale cosmic web filaments \citep{Cantalupo2014,Daddi2021,Tornotti2025}, characterizing large-scale gas kinematics \citep{Zhang2023a,Zhang2023b,Li2021}, and determining CGM metallicities \citep{Guo2020}.
Because the CGM is illuminated by the central ionizing source and part of its kinematics is driven by the galaxy or SMBH itself, these nebulae encode information on physical processes occurring on scales far smaller than the nebulae themselves.
Ly$\alpha$ nebulae can therefore serve as powerful indirect probes of the central engine.

Beyond constraining the CGM, these large-scale nebulae offer a unique window into the structure of high-redshift active galactic nuclei (AGN) on scales of $\leq 100$ pc that remain extremely difficult to resolve directly \citep{GRAVITY2026}.
Although the classical AGN unified model \citep{Antonucci1993,Urry1995,Netzer2015} has received support from direct imaging of the dusty torus \citep{Tristram2014,Carilli2019} and from the comparable space densities of obscured and unobscured AGN in the local Universe \citep{Zakamska2003,Assef2013,Brandt2015}, mounting evidence suggests that this purely orientation-based picture may break down at $z\geq 2$.

Among the most striking discoveries of the James Webb Space Telescope (JWST) are the obscured AGN known as little red dots (LRDs), found at $4\leq z\leq 9$ to be 3–4 orders of magnitude more abundant than their unobscured counterparts at the same epochs \citep{Matthee2024,Kocevski2023,Harikane2023}.
These objects display distinctive ``V-shaped'' spectral energy distributions (SEDs) \citep{Kocevski2025,Setton2025,Hviding2025} and exhibit remarkably weak X-ray emission \citep{Yue2024,Sacchi2025}—properties that are difficult to reconcile with the predictions of the unified scheme \citep{Richards2006,Ross2013}.

While some LRDs have been identified in the local Universe \citep{Lin2026}, these objects are predominantly found at $z=4-9$ \citep{Kocevski2023,Akins2025} and decline rapidly in number density at $z\leq 4$ \citep{Ma2026}, suggesting that they may represent a brief but crucial stage of SMBH growth made possible only in the early Universe through super-Eddington accretion \citep{Inayoshi2025}.
However, recent observations have revealed that obscured quasars at $z=1.0-3.5$ share remarkably similar SEDs with LRDs, indicating that these objects could be the luminous, low-redshift analogues of the high-$z$ LRD population \citep{Wang2025,Wang2025c}.
Probing the small-scale physics of these obscured quasars could therefore hold the key to understanding the nature of high-$z$ LRDs.

Simulations of nebulae around unobscured quasars have demonstrated that the CGM can be used to constrain the properties of the AGN engine, including the SMBH mass and the Eddington ratio \citep{Obreja2024}.
\citep{Zhang2025a} further showed that, under the unified scheme, nebulae around obscured quasars should exhibit more extended surface brightness (SB) profiles and less symmetric morphologies than those around unobscured quasars—providing a potential observational test of AGN orientation models.
Yet, while observations have predominantly targeted unobscured quasars—with more than 300 Ly$\alpha$ nebulae now detected \citep{Cai2019,Arrigoni2019,Farina2019}—only a handful of individual cases \citep{Sanderson2021,Zhang2023a,Zhang2023b} and small samples of $\sim10$ nebulae around obscured quasars have been studied \citep{Gillette2023,Brok2020}.
Exploiting nebulae around different quasar types to probe high-$z$ AGN models thus remains largely unexplored.

In this work, we present a comprehensive analysis of Ly$\alpha$ nebulae around a large sample of quasars at $z=2-3$ observed with the Keck Cosmic Web Imager (KCWI) \citep{Morrissey2018}.
By combining archival data from multiple programs, we construct a sample spanning a wide range of obscuration properties, including 59 unobscured and 26 obscured quasars.
This sample enables, for the first time, a controlled comparison of CGM properties across different quasar populations to test AGN models at cosmic noon.
After carefully comparing the nebular properties of unobscured and obscured quasars, we find that Ly$\alpha$ nebulae around unobscured quasars exhibit more extended SB profiles and less symmetric morphologies than those around obscured quasars—a result that is opposite to the prediction of the AGN unified scheme \citep{Brok2020,Zhang2025a}.
However, this finding can be naturally incorporated into an evolutionary scenario \citep{Sanders1988,Hopkins2008,Alexander2012}, in which the AGN transitions from an obscured to an unobscured phase following a large-scale feedback event.

This paper is organized as follows.
In Section \ref{sec:data}, we describe the sample selection, data reduction, and derived quasar properties.
Section \ref{sec:results} presents the measurements of nebular morphology, surface brightness, and kinematics.
In Section \ref{sec:discussion}, we investigate the connection between CGM properties and quasar characteristics, and compare our results with predictions from the AGN unified model and evolutionary scenarios.
Finally, we summarize our findings and discuss their implications for AGN–CGM co-evolution in Section \ref{sec:conclusion}. 
All wavelengths and line names are given in the vacuum frame. 
All equivalent widths (EWs) are reported in the rest frame. 
A flat $\Lambda$CDM cosmology is assumed, with $H_{\rm 0}=70$ km s$^{-1}$ Mpc$^{-1}$, $\Omega_{\Lambda, 0}=0.7$ and $\Omega_{m, 0}=0.3$.

\section{Data} \label{sec:data}

\subsection{Sample selection} \label{sec:selection}
Datacubes of 96 quasars at $z=2-3$ obtained during 2017-2022 are collected from the KCWI public archive. 
These data span 11 separate programs. 
Since four of the programs target specific circumgalactic environments and thus introduce strong selection biases, we restrict our analysis to the remaining seven programs, yielding a final sample of 85 quasars.
The selection criteria of each program are summarized below.

{\bf (i) Following the Rise and Fall of Enormous Ly$\alpha$ Nebulae with KCWI (PI: J. Xavier Prochaska \& Zheng Cai).}
This program comprises 27 quasars. 
By cross-matching with the Sloan Digital Sky Survey (SDSS) DR16 catalog, we identify six quasars as obscured narrow-line (NL) quasars based on the criterion $\rm FWHM_{\rm CIV}\leq 2000$ km s$^{-1}$ \citep{alexandroff2013,Law2018}. 
One of these resides in the MAMMOTH-1 overdensity field \citep{Cai2017}; the remaining five are drawn from an obscured NL candidate catalog \citep{alexandroff2013}, in which candidates were selected from the Baryon Oscillation Spectroscopic Survey (BOSS) of SDSS-III on the basis of weak optical continua together with strong Ly$\alpha$ and \civ\ emission.
These quasars exhibit dust extinctions of $A_{\rm V}\approx 0.5$ mag, indicating modest obscuration.
The remaining 21 broad-line quasars were selected by restricting their redshifts to $2.1\leq z\leq2.3$ and their $i$-band magnitudes to $i\leq 18.5$ \citep{Cai2019}.

{\bf (ii) The FLASHES Survey - Witnessing Galaxy Formation (PI: Christopher Martin). }
This program contributes 20 quasars drawn from the Fluorescent Lyman-Alpha Structures in High-$z$ Environments Survey (FLASHES) \citep{OSullivan2020}.
The survey targets quasars in the redshift range $2.3\leq z\leq 3.1$.
To enable an investigation of the luminosity dependence of nebular properties, the sample was designed to span a wide range of absolute $i$-band magnitudes.	
Where available, radio-quiet (RQ) sources were preferentially selected.
Among the 20 quasars, four satisfy the obscured quasar criterion $W_{2}-W_{4}\geq 4.8$ \citep{OSullivan2020}; two of these four are further classified as NL quasars.

{\bf (iii) Mapping the CGM of Forming Galaxies (PI: Charles C. Steidel).} 
This program includes 10 quasars, five of which are classified as obscured.
Four of the five obscured AGN were initially identified as star-forming galaxies (SFGs) in the Keck Baryonic Structure Survey (KBSS) fields but were subsequently reclassified as AGN based on their positions on the BPT diagram.
Spectroscopy reveals $\rm FWHM_{\rm CIV}\leq 1300$ km s$^{-1}$ for three of these sources \citep{Law2018}, confirming their NL nature, and SED fitting with obscured AGN templates further supports their classification as obscured objects \citep{Law2018}.
The fourth source shows $\rm FWHM_{\rm H\alpha}\leq 600$ km s$^{-1}$ \citep{Steidel2014}, which is also consistent with an NL quasar; we therefore classify it as an obscured quasar on the basis of its selection criteria and spectral features.
The fifth obscured quasar was identified through X-ray observations that reveal a hydrogen column density of $N_{\rm H}\geq 10^{23}$ cm$^{-2}$ \citep{Geach2009}.
The remaining five quasars are unobscured: three were selected by their redshifts ($2.25\leq z\leq 2.27$) and brightness ($i\leq 18.5$) \citep{Langen2023}, while the other two lie in the KBSS field.

{\bf (iv) Mapping the Extended Infall/Outflow Gas Around Extremely Red Quasars (PI: Fred Hamann).}
This program contains 10 quasars identified as obscured broad-line (BL) quasars—so-called red quasars—selected from BOSS by applying a color cut of $i-W_{3}\geq 4.6$ \citep{Hamann2017}.
The photometry of these sources is well described by a reddened quasar template with $\rm E(B-V)\approx0.3$, confirming their dust-obscured nature \citep{Hamann2017}.
[\oiii] line measurements are available for five of the 10 quasars and indicate the presence of potential outflows \citep{Gillette2023}.

{\bf (v) Resolving Distant Quasar Host Galaxies and Their Environments (PI: Shelley A. Wright).}
This program includes six BL quasars at $2.0\leq z\leq 2.6$.
Four were selected from SDSS DR10 on the basis of $R\leq 17.5$ and a 1.4 GHz radio flux $\geq 0.3$ mJy \citep{Vayner2019}; the remaining two, although not observed by SDSS, satisfy the same criteria \citep{Vayner2019}.
The sample is therefore biased toward radio-loud (RL) quasars.
Notably, one of the six quasars has $r-W_{4}\approx 5.6$ AB mag, a value comparable to those of the obscured quasars in our sample.
SED fitting strongly supports an obscured classification for this source, and we accordingly treat it as an obscured BL quasar.
\begin{figure}
\centering
\includegraphics[width=\columnwidth]{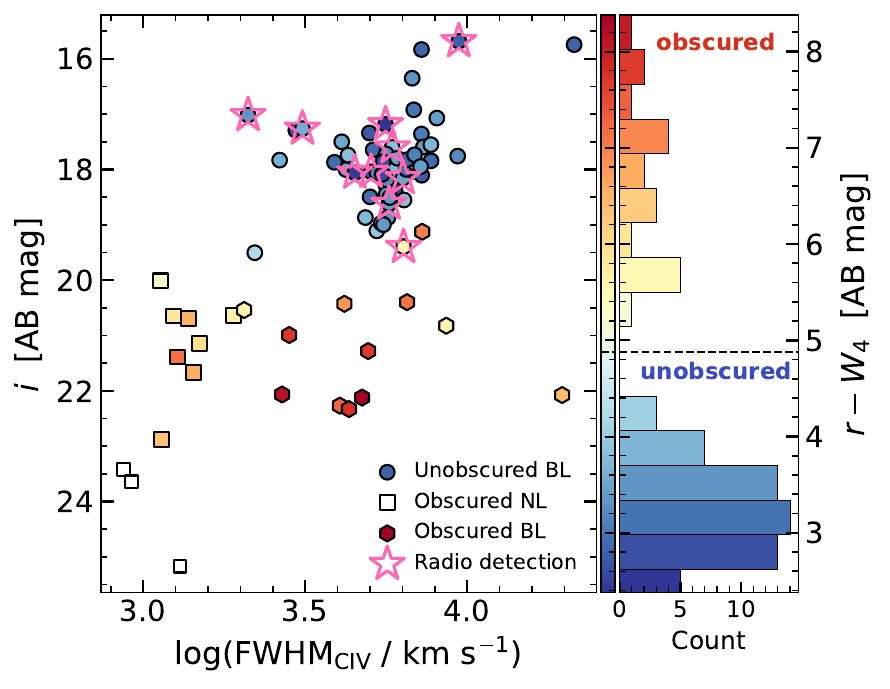}
\caption{Properties of the quasars in our sample. 
The $i$-band magnitude is plotted against the \civ\ emission-line width, with different markers indicating different quasar types. 
Pink stars mark radio-detected quasars (excluding three without SDSS DR16 matches). 
The color coding represents $r-W_{4}$, which traces the level of quasar obscuration. 
The sample separates into two distinct populations at $r-W_{4}=5$ AB mag (horizontal dashed line), yielding 26 obscured and 59 unobscured quasars.}
\label{sample_properties}
\end{figure}

{\bf  (vi) A comprehensive study of the CGM at $z\sim 2$ (PI: Kathy L. Cooksey).}
This program contains seven BL quasars drawn from the Advanced Camera for Surveys + Wide Field Camera 3 Survey \citep{OMeara2013}.
The quasars were selected to have redshifts in the range $2.3\leq z\leq 2.6$ and $g$-band magnitudes of $g\leq 18.5$, with the additional requirement that their spectra show no strong associated broad absorption lines (BALs).

{\bf (vii) The Physics of Gas Flows Around Galaxies at Cosmic Noon (PI: Glenn G. Kacprzak).}
This program contributes five BL quasars, originally targeted to study foreground \mgii\ absorber—galaxy pairs at $1.9\leq z\leq 2.6$ \citep{Nielsen2020}.
Aside from the exclusion of background quasars with associated \mgii\ and \civ\ absorbers, no additional selection criteria were imposed.

The above description indicates that most of the quasars in our sample are selected by their magnitudes, colors, and redshift ranges. 
Applying the criteria of \cite{alexandroff2013} and \cite{Law2018}, we identify 13 of the 85 quasars as narrow-line (NL) quasars; the remainder are broad-line (BL) quasars.
Adopting the color cut $r-W_{4}\geq 5$ AB mag to identify dust-obscured objects at high redshift \citep{Hamann2017,Wang2025,Wang2025b}, we find that 26 quasars (including all 13 NL quasars) satisfy this criterion and are classified as obscured, while the remaining 59 are classified as unobscured.
We also cross-match the sample against the Faint Images of the Radio Sky at Twenty-Centimeters (FIRST) and the Very Large Array Sky Survey (VLASS) catalogs: 13 quasars are detected in the radio band, 57 remain undetected, and 15 fall outside the survey footprints.
These properties are summarized in Fig.~\ref{sample_properties} and Table~\ref{quasar_information}.
Missing values in the table indicate quantities that were not measured for a given source.

{\setlength{\LTcapwidth}{6.5in}
\begin{longtable*}{p{2.5cm}p{1cm}p{1.1cm}p{1.1cm}p{1.2cm}p{1cm}p{1cm}}
\caption{{\bf  Quasar Properties at $z\approx 2$ in our sample.} Col. (1): Source name. 
Col. (2): Redshift estimated from the Ly$\alpha$ emission-line peak within a $1''$ aperture. 
Col. (3): Right ascension from SDSS DR16 or NED.
Col. (4): Declination from SDSS DR16 or NED.
Col. (5): \civ\ emission-line FWHM from SDSS DR16 or previous observations \citep{Steidel2014,Cai2017,Law2018}.
Col. (6): $r-W_{4}$ color index in AB mag.
Col. (7): Program number (Roman numeral).}\\
\hline
\hline
          Name$^{(1)}$ & $z_{\rm Ly\alpha, peak}^{(2)}$ &          \multicolumn{1}{p{2cm}}{\centering RA$^{(3)}$ \\ (deg)} &         \multicolumn{1}{p{2cm}}{\centering DEC$^{(4)}$ \\ (deg)}& \multicolumn{1}{p{2cm}}{\centering FWHM$_{\rm CIV}^{(5)}$ \\ (km s$^{-1}$)} & $r-W_{4}$ (AB mag) &  \multicolumn{1}{p{1cm}}{\centering Sample} \\
\hline
\endfirsthead
\hline
\hline
          Name$^{(1)}$ & $z_{\rm Ly\alpha, peak}^{(2)}$ &          \multicolumn{1}{p{2cm}}{\centering RA$^{(3)}$ \\ (deg)} &         \multicolumn{1}{p{2cm}}{\centering DEC$^{(4)}$ \\ (deg)}& \multicolumn{1}{p{2cm}}{\centering FWHM$_{\rm CIV}^{(5)}$ \\ (km s$^{-1}$)} & $r-W_{4}$ (AB mag) &  \multicolumn{1}{p{1cm}}{\centering Sample} \\
\hline
\endhead
\hline
\multicolumn{7}{c}{{{\bf Table 1} (Continued)}} \\
\endfoot

\endlastfoot
1227+2848 & 2.3 & 186.865 & 28.813 & 6880 & 3.1 & i \\
1228+3128 & 2.2 & 187.104 & 31.477 & 9424 & 2.8 & i \\
1230+3320 & 2.3 & 187.648 & 33.333 & 9347 & 3.2 & i \\
1416+2649 & 2.3 & 214.072 & 26.818 & 5049 & 2.7 & i \\
1426+2555 & 2.3 & 216.649 & 25.923 & 7327 & 3.4 & i \\
1441+4003 & 2.3 & 220.351 & 40.053 & 1140 & 6.4 & i \\
1444+3904 & 2.3 & 221.233 & 39.067 & 6000 & 2.5 & i \\
1517+0055 & 2.6 & 229.446 & 0.931 & 1132 & 5.3 & i \\
1518+0239 & 2.5 & 229.738 & 2.657 & 7183 & 3.4 & i \\
 J0011 & 2.3 & 2.927 & -0.146 & 1241 & 5.8 & i \\
 J0046 & 2.5 & 11.502 & 0.095 & 1278 & 7.2 & i \\
 J1020+0259\_off & 2.5 & 155.153 & 3.029 & 1897 & 5.6 & i \\
 J1035-0234\_off & 2.3 & 158.992 & -2.556 & 1492 & 6.0 & i \\
 J1432 & 2.8 & 236.590 & 2.017 & 6335 & 3.5 & i \\
 J2123.f & 2.3 & 320.873 & -0.848 & 6766 & 3.3 & i \\
 Q0048 & 2.3 & 12.235 & 0.947 & 6486 & 2.9 & i \\
 Q0050 & 2.2 & 12.588 & 0.860 & 2637 & 3.6 & i \\
 Q0052 & 2.3 & 13.140 & 1.678 & 7224 & 3.0 & i \\
 Q0101 & 2.4 & 15.319 & 2.033 & 5656 & 3.2 & i \\
 Q0107 & 2.3 & 16.904 & 3.250 & 6840 & 3.0 & i \\
 Q0848 & 2.3 & 132.237 & -1.250 & 5607 & 3.3 & i \\
 Q2121 & 2.4 & 320.496 & 0.873 & 5468 & 2.6 & i \\
 Q2127 & 2.3 & 321.948 & 0.825 & 7727 & 3.7 & i \\
 QG0814 & 2.2 & 123.506 & 32.847 & 5771 & 3.4 & i \\
 QG0814\_QSOoff & 2.4 & 124.515 & 32.198 & 5694 & 3.3 & i \\
 QSOoffsetCorr & 3.0 & 353.639 & 23.600 & 6381 & 3.8 & i \\
 jackpot\_QSOoff & 2.2 & 130.650 & 38.341 & 5431 & 4.3 & i \\
 HS1700-1.f & 2.7 & 255.253 & 64.203 & 7228 & 2.8 & ii \\
 J1705.1 & 2.4 & 256.494 & 27.607 & 1377 & 6.5 & ii \\
 RL\_AGN & 2.8 & 154.358 & 61.274 & 4501 & 2.4 & ii \\
 SDSS0006+1614A & 2.4 & 1.664 & 16.250 & 7240 & 2.8 & ii \\
 SDSS0041+1925 & 2.7 & 10.291 & 19.422 & 2210 & 4.2 & ii \\
 SDSS0132+3326 & 2.4 & 23.186 & 33.449 & 5261 & 3.9 & ii \\
 SDSS0137\_2405 & 2.4 & 24.494 & 24.095 & 1429 & 6.6 & ii \\
 SDSS0205+1902A & 2.7 & 31.365 & 19.041 & 8061 & 3.4 & ii \\
 SDSS0211\_3117 & 2.8 & 32.914 & 31.290 & 4851 & 3.7 & ii \\
 SDSS0214\_1912 & 2.5 & 33.624 & 19.210 & 4290 & 3.8 & ii \\
 SDSS0300+0222 & 2.5 & 45.192 & 2.379 &  & 3.4 & ii \\
 SDSS0730+4340D & 3.0 & 112.512 & 43.668 & 5515 & 3.4 & ii \\
 SDSS0834+1238.4 & 2.7 & 128.536 & 12.643 & 5977 & 3.7 & ii \\
 SDSS0837+1459 & 2.5 & 129.304 & 14.988 & 5628 & 3.3 & ii \\
 SDSS0958.f & 2.5 & 149.689 & 47.057 & 5126 & 2.9 & ii \\
 SDSS1112+1521 & 2.8 & 168.219 & 15.357 & 6508 & 2.8 & ii \\
 SDSS1428+2336 & 2.8 & 217.046 & 23.611 & 5628 & 3.3 & ii \\
 SDSS2241\_1225 & 2.7 & 340.438 & 12.433 & 7736 & 2.9 & ii \\
 SDSS2340+2418 & 2.3 & 355.166 & 24.316 & 2050 & 5.6 & ii \\
 SDSS2350+3135A & 2.8 & 357.652 & 31.585 & 8619 & 5.6 & ii \\
 J0010+0617 & 2.3 & 2.524 & 6.289 & 5016 & 2.9 & iii \\
 J0030+0530 & 2.3 & 7.591 & 5.515 & 5265 & 3.1 & iii \\
 J2316+0906 & 2.3 & 349.206 & 9.110 & 4228 & 3.5 & iii \\
 Q0100-BX170 & 2.3 & 15.785 & 13.278 & 870 &  & iii \\
 Q0100.2 & 2.7 & 15.801 & 13.273 &  &  & iii \\
 Q0142-BX.f & 2.4 & 26.324 & -9.748 & 920 &  & iii \\
 Q0207-BX298 & 2.1 & 32.475 & -0.075 & 1300 &  & iii \\
 Q1549-BX101 & 2.4 & 237.980 & 19.159 &  &  & iii \\
 Q1549\_QSO\_E & 2.9 & 237.969 & 19.184 & 21387 & 2.8 & iii \\
 SSA22\_Blob2 & 3.1 & 334.413 & 0.225 &  &  & iii \\
 J0006 & 2.3 & 1.544 & 12.250 & 4737 & 8.9 & iv \\
 J0220 & 3.1 & 35.217 & 1.620 & 2683 & 8.1 & iv \\
 J083448+015921 & 2.6 & 128.702 & 1.989 & 2823 & 7.7 & iv \\
 J114508+574258 & 2.9 & 176.283 & 57.716 & 7266 & 7.0 & iv \\
 J123241+091209 & 2.4 & 188.174 & 9.203 & 4949 & 7.7 & iv \\
 J145113+013234 & 2.8 & 222.807 & 1.543 & 6522 & 7.1 & iv \\
 J145148+233845 & 2.6 & 222.950 & 23.646 & 4180 & 6.8 & iv \\
 J2215 & 2.5 & 333.850 & -0.946 & 4318 & 7.8 & iv \\
 J2254+2327 & 3.1 & 343.660 & 23.454 & 19635 & 6.5 & iv \\
 J2323 & 2.4 & 350.859 & -1.009 & 4044 & 7.3 & iv \\
 3C9 & 2.0 & 5.105 & 15.682 & 5599 & 2.3 & v \\
 4C04 & 2.6 & 355.242 & 4.521 & 6362 & 5.5 & v \\
 4C05 & 2.3 & 336.311 & 5.453 & 5868 & 3.8 & v \\
 4C09 & 2.1 & 72.091 & 9.848 &  &  & v \\
 4C11 & 2.2 & 200.328 & 11.114 & 5737 & 3.4 & v \\
 4C57 & 2.2 & 254.941 & 57.525 &  &  & v \\
 J0340-0519 & 2.3 & 55.102 & -5.319 & 3894 & 2.9 & vi \\
 J0755+2204 & 2.3 & 118.949 & 22.081 & 5201 & 3.0 & vi \\
 J1143+0524 & 2.6 & 175.994 & 5.413 & 6841 & 3.0 & vi \\
 J1325+6634 & 2.5 & 201.467 & 66.568 & 5496 & 2.6 & vi \\
 J1540+4138 & 2.5 & 235.179 & 41.615 & 4992 & 2.7 & vi \\
 J1651+4002 & 2.3 & 252.906 & 40.039 & 3102 & 3.6 & vi \\
 J1724 & 2.6 & 261.038 & 53.235 & 6085 & 3.4 & vi \\
 J115944+011206 & 2.0 & 179.937 & 1.202 & 2107 & 3.3 & vii \\
 J122607+173650 & 3.0 & 186.530 & 17.614 & 6976 & 4.2 & vii \\
 J143040+014939 & 2.1 & 217.670 & 1.828 & 4100 & 3.5 & vii \\
 J223408 & 3.0 & 338.537 & 0.000 & 2959 & 2.7 & vii \\
 QJ2343.f & 2.6 & 356.618 & 12.817 &  &  & vii \\
\hline
\hline
\label{quasar_information}
\end{longtable*}}

\subsection{Sample properties} \label{data_sample_phys_properties}
We cross-match these quasars with the SDSS DR16, the Wide-field Infrared Survey Explorer (WISE), and the NASA/IPAC Extragalactic Database (NED) to obtain their photometry. 
We compile photometry spanning $0.4$–$22.8\,\mu\mathrm{m}$ from these databases for SED fitting, a wavelength range that captures continuum emission from the AGN accretion disc, the torus, and the host galaxy.

We use \textsc{cigale} \citep{Boquien2019} to fit the SEDs, adopting the parameter settings from \citet{Zhang2023c}, who applied \textsc{cigale} to AGN SEDs at $z\approx 2$.
We fix the dust torus opening angle at 50 deg, consistent with typical constraints on AGN torus geometry \citep{Fritz2006}.
Because the \citet{Zhang2023c} sample includes only unobscured quasars, a viewing angle of 90 deg between the equatorial axis and the line of sight is assumed—that is, a configuration in which the torus does not obscure the sightline.
For the obscured quasars in our sample, we instead set this angle to 0 deg, corresponding to a sightline passing through the torus. 
We further constrain the galaxy stellar mass to $\leq 10^{12} \ M_{\odot}$ to avoid the unphysical results. 

\begin{figure*}
\centering
\includegraphics[width=\textwidth]{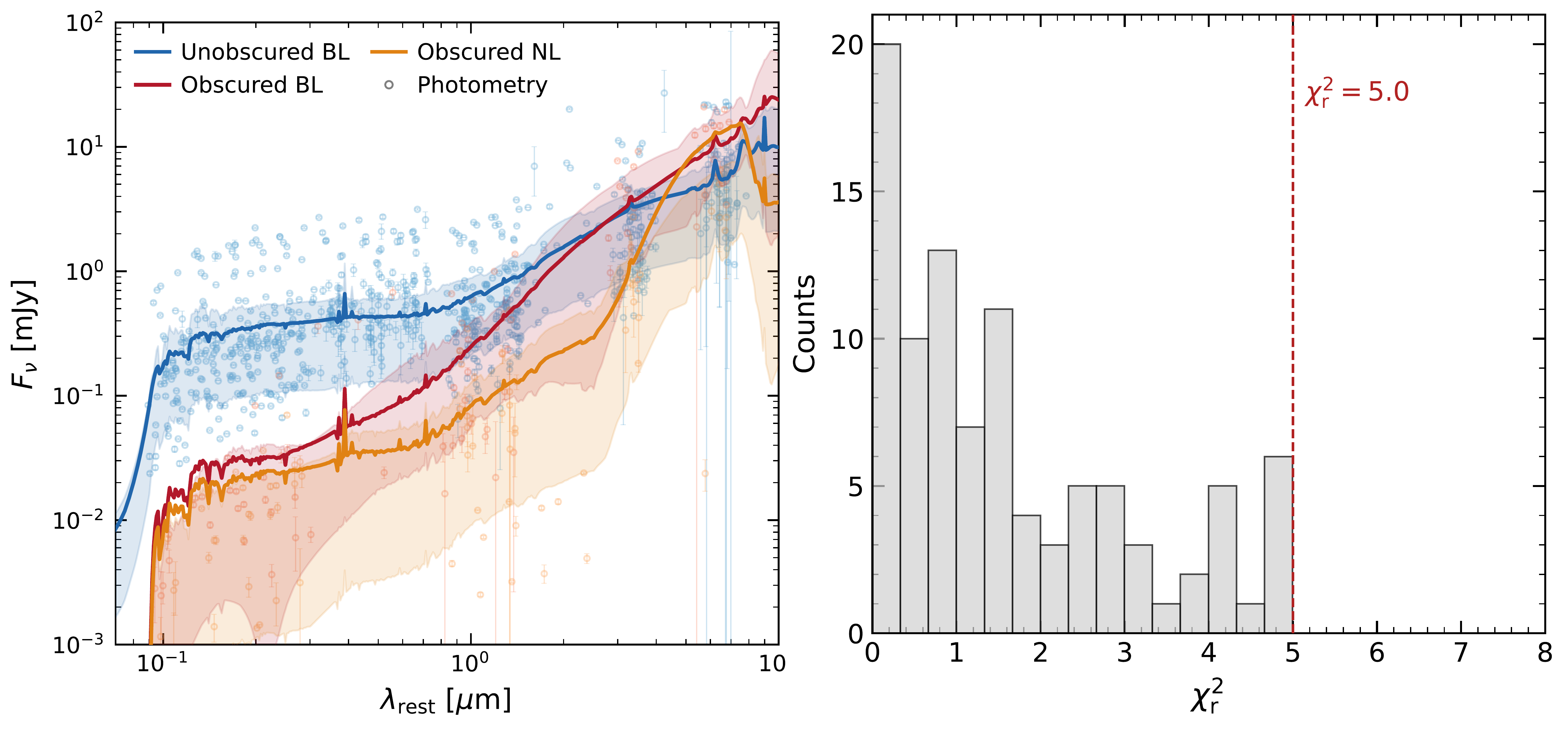}
\caption{{\bf Left:} The rest-frame photometries (empty circles) and best-fit SEDs (solid lines) of quasars. 
The shaded regions marking the 16th-84th percentile range. 
{\bf Right:} Reduced chi-square ($\chi_{r}^{2}$) from the SED fitting; the red dashed line marks the acceptable threshold of $\chi_r^{2} \leq 5$.
The resulting SED fits further support the obscured–unobscured classification based on the $r-W_{4}$ color criteria.}
\label{qso_sed}
\end{figure*}

Given that previous observations yield good fits with a reduced chi-square of $\chi^{2}_{\rm r}\leq 5$ \citep{Zhang2023c}, we adopt this threshold as the upper acceptable limit.
The best-fit rest-frame SEDs and the $\chi^{2}_{\rm r}$ distribution are shown in Fig.~\ref{qso_sed}.
At $\chi^{2}_{\rm r}\leq 5$, the observed photometry falls within the 1-$\sigma$ scatter of the model, indicating satisfactory fits.
We note that we fit each obscured quasar under both unobscured and obscured assumptions; the obscured scenario consistently yields a lower $\chi^{2}_{\rm r}$, supporting their classification as obscured objects.

The stellar mass ($M_{\star}$), AGN bolometric luminosity ($L_{\rm bol}$), and $\chi^{2}_{\rm r}$ are obtained from the fitting process. 
The AGN bolometric luminosity reported by \textsc{cigale} is the sum of the disk emission, dust re-emission, and scattered emission \citep{Boquien2019}, which we use to estimate $M_{\rm BH}$ under the Eddington limit via Eq.~\ref{MBH_Eddington_limit}.
\begin{equation}
    M_{\rm BH}=\frac{L_{\rm bol}}{1.26\times 10^{38}} \ {\rm M_{\odot}}
    \label{MBH_Eddington_limit}
\end{equation}

\begin{figure}
\centering
\includegraphics[width=\columnwidth]{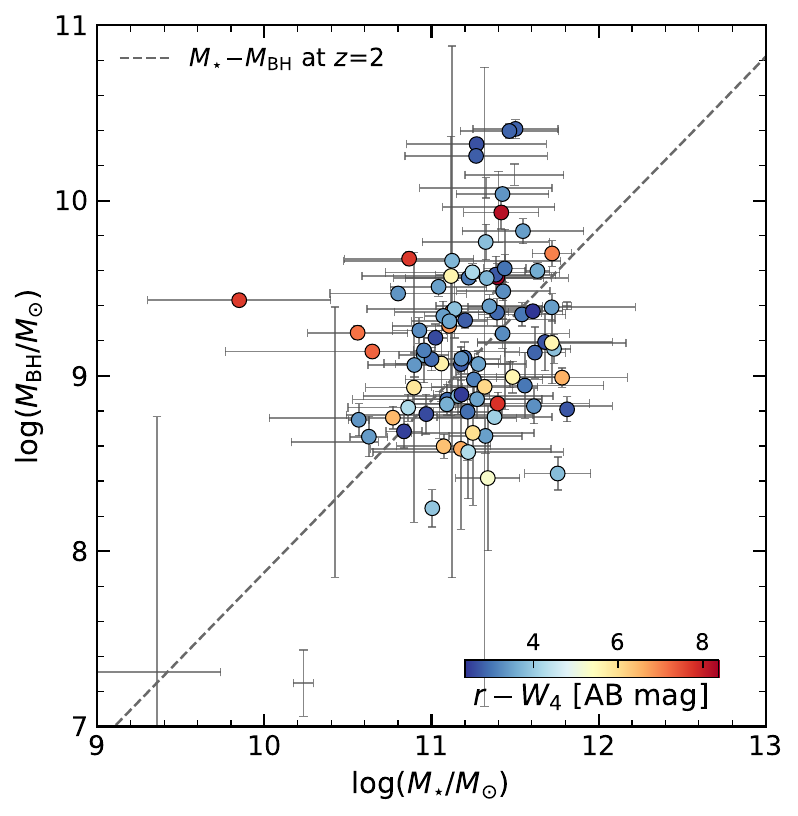}
\caption{The host galaxy stellar mass versus the SMBH mass. 
The stellar mass is derived from the SED fitting, while the SMBH mass is estimated using Eq.~\ref{MBH_Eddington_limit} with the quasar bolometric luminosity from the SED fitting. 
The dot color denotes the $r-W_{4}$. 
The dashed line shows the best linear fit to the $M_{\star}$-$M_{\rm BH}$ relation from previous observations on AGNs at $z=2-3$ \citep{Zhang2023c}. }
\label{mbh_vs_ms}
\end{figure}

The resulting stellar and SMBH masses are presented in Fig.~\ref{mbh_vs_ms}.
The median stellar masses for the three quasar populations are $M_{\star}=10^{11.2\pm0.3}\,M_{\odot}$ (unobscured BL), $10^{11.1\pm0.5}\,M_{\odot}$ (obscured BL), and $10^{11.1\pm0.6}\,M_{\odot}$ (obscured NL), while the median SMBH masses are $M_{\rm BH}=10^{9.2\pm0.4}\,M_{\odot}$ (unobscured BL), $10^{9.3\pm0.3}\,M_{\odot}$ (obscured BL), and $10^{8.6\pm0.6}\,M_{\odot}$ (obscured NL). 
The three populations share broadly consistent stellar and SMBH masses within the 1-$\sigma$ scatter, although $M_{\rm BH}$ of the obscured NL quasars is slightly lower. 
The fitting yields an $M_{\star}$–$M_{\rm BH}$ relation consistent with previous observations of AGNs at $z\approx 2$ \citep{Zhang2023c}, further confirming the overall reliability of the SED fits. 

We also explore the radio properties of these quasars. 
We calculate the radio loudness $R_{K}$ using Eq.~\ref{RK_eq}, following \citet{Kellermann1989,Mackenzie2021}:
\begin{equation}
    R_{K}=\frac{f_{\rm \nu, 5GHz}}{f_{\rm \nu, 4400A}}
    \label{RK_eq}
\end{equation}
where $f_{\rm \nu, 5GHz}$ and $f_{\rm \nu, 4400A}$ are the rest-frame flux densities. 
For quasars at $z=2-3$, these correspond to observed frequencies and wavelengths of $1.3-1.7$ GHz and $1.3-1.8\ \mu$m, respectively.
Because this frequency range is centered near $1.5$ GHz—close to the $1.4$ GHz center frequency of the FIRST survey—we cross-match our sample against the FIRST catalog to obtain the radio flux densities. 
Only 11 quasars are detected in FIRST; 57 remain undetected and 17 fall outside the survey footprint.
We also cross-match the sample with the VLASS catalog at 3 GHz, recovering two additional detections (the 67th and 70th entries in Tab.~\ref{quasar_information}). 
To estimate the 1.4 GHz flux densities of these two VLASS-detected sources, we apply the median spectral scaling factor derived from quasars detected in both FIRST and VLASS.

For detected sources, we classify quasars as radio-loud (RL) or radio-quiet (RQ) using the threshold $R_{\rm K}=10$.
For undetected sources, we compute $R_{\rm K}$ from the 3-$\sigma$ upper limit on the radio flux.
Sources whose upper limit yields $R_{\rm K}<10$ are classified as RQ (with $R_{\rm K}$ fixed at 5), while those exceeding this threshold are deemed ambiguous (with $R_{\rm K}$ fixed at 10).
In total, 12 quasars are classified as RL, 56 as RQ, and 17 as ambiguous.


\subsection{Data reduction} \label{sec:data_reduction}
The standard KCWI pipeline \footnote{https://github.com/Keck-DataReductionPipelines/KcwiDRP} is employed to reduce the data and produce the datacubes. 
We subtract the bias, correct the pixel-to-pixel variation with the flat-field images, and remove the cosmic rays. 
Then, the geometry transformation and the wavelength calibration with the ThAr arc are performed. 
The twilight flat images are used to correct the slice-to-slice variance. 
The spectrophotometric standard stars are adopted for calibrating the flux of the individual image in the datacube. 
The sky subtraction is also done by scaling the flux as a function of wavelength in the sky datacube to match the data. 

The \textsc{cwitools} \citep{OSullivan2020b} is applied for coadding the individual datacube. 
The built-in method, \texttt{cwi\_coadd}, calculates the footprint for each pixel on the coadd frame and distributes its flux onto this frame. 
The new world coordinate system (WCS) for the coadd frame is constructed to achieve a 1:1 aspect ratio, resulting in a pixel scale of $0.3''\times 0.3''$. 
The point spread function (PSF) is subtracted for the unobscured BL quasars, as their bright PSFs contaminate the faint nebular emission. 
This is carried out by the \texttt{cwi\_psf\_sub} function from the \textsc{cwitools}. 
Our sample comprises previously observed nebulae \citep{Cai2019}, with consistent SB profiles and morphologies matching prior results, validating the data reduction process. 
We obtain the final stacked datacubes after reduction. 
The median depth at 4250 \AA \ over a width of 1 \AA \ under an aperture of $1''$ for the three types of quasar samples are $\sigma_{\rm SB} ({\rm unobs\ BL})=4.2\times 10^{-19}$ erg s$^{-1}$ cm$^{-2}$ arcsec$^{-2}$, $\sigma_{\rm SB} ({\rm obs\ BL})=4.0\times 10^{-19}$ erg s$^{-1}$ cm$^{-2}$ arcsec$^{-2}$, and $\sigma_{\rm SB} ({\rm obs\ NL})=4.2\times 10^{-19}$ erg s$^{-1}$ cm$^{-2}$ arcsec$^{-2}$, respectively. 
The median seeing are $\approx1''$.


\section{Results} \label{sec:results}
\begin{figure*}
\centering
\includegraphics[width=\textwidth]{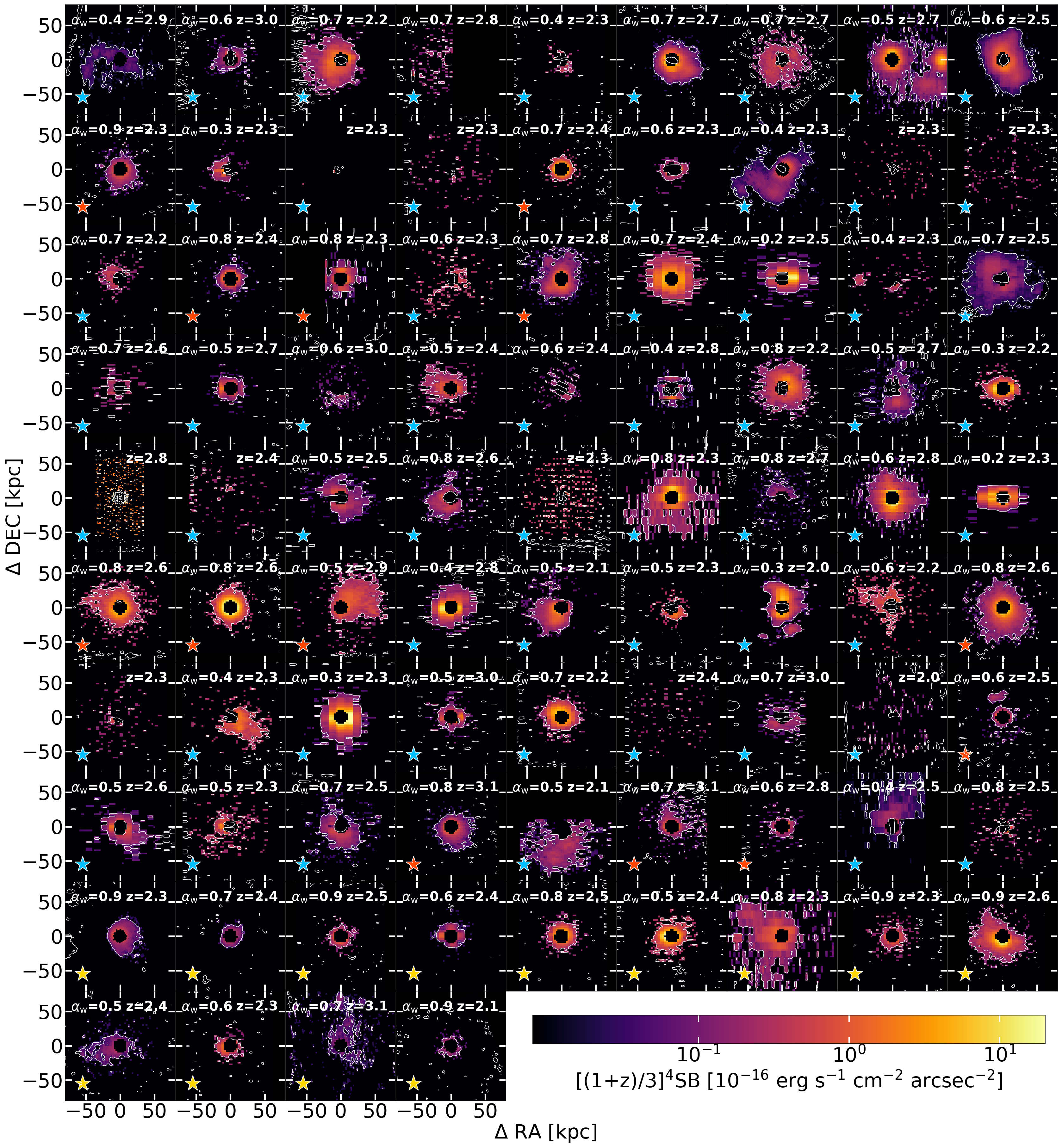}
\caption{The NB images of Ly$\alpha$ nebulae. 
We use a window centering on the Ly$\alpha$ line with a spectral width of $\pm 1500$ km s$^{-1}$ to extract the NB images. 
A central region within 15 kpc ($\approx 1.8''$ at $z=2-3$) is masked to avoid PSF contamination. 
The white contour represents the 2-$\sigma$ SB noise. 
Each panel displays the $\alpha_{\rm w}$ and redshift. 
The missing $\alpha_{\rm w}$ values means non-detection of the nebulae. 
The quasar type is indicated in the lower left by blue (unobscured BL), red (obscured BL), and orange (obscured NL) stars.}
\label{nb_img}
\end{figure*}

We construct pseudo-narrowband (pseudo-NB) images centered on the Ly$\alpha$ emission line with a spectral width of $\pm1500$ km s$^{-1}$ and adopt a signal-to-noise threshold of $S/N\geq 2$ to define the emitting regions \citep{Cai2019}.
A circular region of 15 kpc ($\approx 1.8''$) around each quasar is masked to exclude PSF-dominated emission.
The resulting NB images are presented in Fig.~\ref{nb_img}.
For nebulae previously observed by \citet{Cai2019,Zhang2023a,Zhang2023b}, we recover consistent SB profiles and morphologies, confirming the reliability of our reduction.

We examine the nebular morphology (Sec.~\ref{nebular_morphology}), surface brightness profiles (Sec.~\ref{nebular_SB}), and kinematics (Sec.~\ref{nebular_velocity_dispersion}) in detail.
Our observations reveal notable differences from the predictions of the unified model simulations, which we discuss in Sec.~\ref{sec:discussion}.

\subsection{Nebular morphology} \label{nebular_morphology}
\begin{figure*}
\centering
\includegraphics[width=\textwidth]{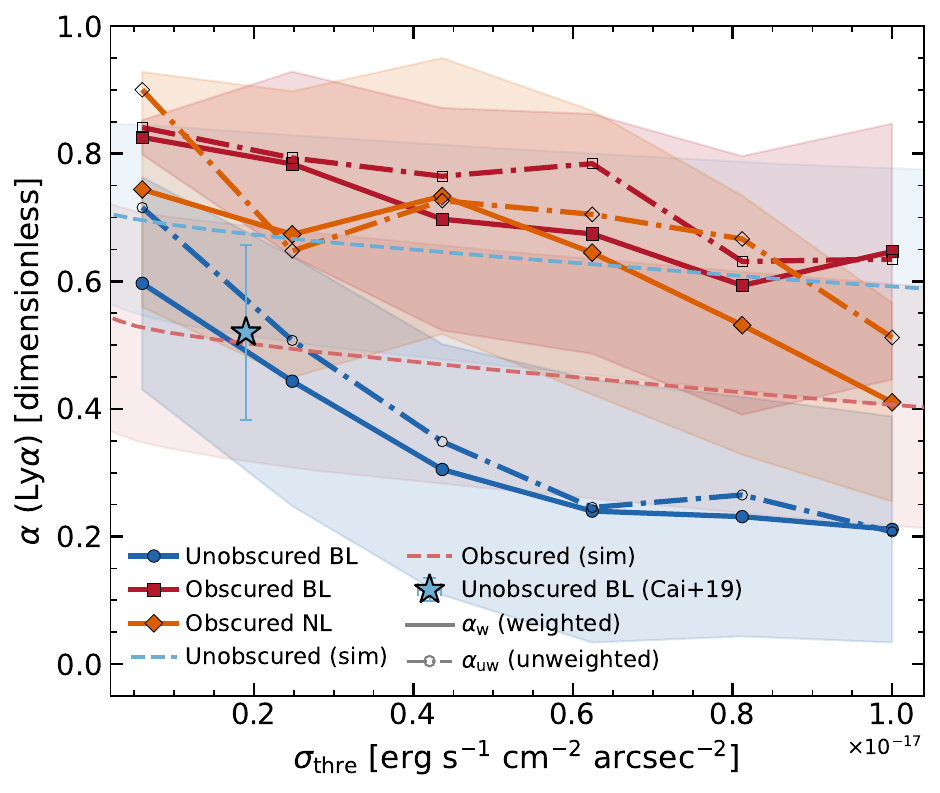}
\caption{Stacked asymmetry profiles for Ly$\alpha$ nebulae around unobscured (blue) and obscured (red and orange) quasars.
The solid and dash-dotted lines show the flux-weighted and unweighted $\alpha$ parameters, respectively.
The dashed line represents the simulated profile from the unified model \citep{Zhang2025a}, and the blue star marks previous observations \citep{Cai2019}.
Shaded regions denote the 1-$\sigma$ scatter.
Nebulae around obscured quasars exhibit larger $\alpha$ values, indicating more symmetric morphologies—a trend opposite to the predictions of the unified model.}
\label{alpha_vs_sigma}
\end{figure*}

We adopt the method of \cite{Arrigoni2019,Cai2019,Herwig2024} to calculate the $\alpha_{\rm w}$ which quantifies the symmetry of the Ly$\alpha$ nebulae (Eq.~\ref{moment_for_alpha}). 
\begin{subequations}
\label{moment_for_alpha}
    \begin{align}
        & M_{xx}=\langle \frac{(x-x_{\rm AGN})^{2}}{r^{2}} \rangle_{f} \\
        & M_{yy}=\langle \frac{(y-y_{\rm AGN})^{2}}{r^{2}} \rangle_{f} \\
        & M_{xy}=\langle \frac{(x-x_{\rm AGN})(y-y_{\rm AGN})}{r^{2}} \rangle_{f} \\
        & \alpha_{\rm w}=\frac{1-\sqrt{(M_{xx}-M_{yy})^{2}+(2M_{xy})^{2}}}{1+\sqrt{(M_{xx}-M_{yy})^{2}+(2M_{xy})^{2}}}
    \end{align}
\end{subequations}
Here $(x,y)$ denotes the pixel position, $(x_{\rm QSO}, y_{\rm QSO})$ is the quasar position, and $r$ is their separation.
The second-order moments $M_{xx}$, $M_{yy}$, and $M_{xy}$ are computed using the pixel flux as a weight.
Pixels within $1.8''$ of the quasar are excluded from the calculation. 
The parameter $\alpha_{\rm w}$ ranges from 0 to 1, with $\alpha_{\rm w}=1$ corresponding to perfect circular symmetry.
For each datacube, we vary the SB threshold from 2-$\sigma_{\rm SB}$ to 200-$\sigma_{\rm SB}$ in 50 steps to define the emitting region over which $\alpha_{\rm w}$ is computed. 
The resulting $\alpha_{\rm w}$ profiles are then stacked by quasar type, as shown in Fig.~\ref{alpha_vs_sigma}. 

Over the SB range $2.0\times 10^{-19}-1.0\times 10^{-17}$ erg s$^{-1}$ cm$^{-2}$ arcsec$^{-2}$, nebulae around unobscured quasars exhibit $\alpha_{\rm w, unobs}=0.2-0.6$ (Fig.~\ref{alpha_vs_sigma}), consistent with previous observations \citep{Cai2019}.
In contrast, nebulae around obscured quasars display larger values ($\alpha_{\rm w, obs}=0.6-0.8$), indicating more symmetric morphologies. 
A paired t-test on these $\alpha_{\rm w}$ profiles yields $p\approx 0.77\%$, confirming that the difference is statistically significant at the 5\% level. 
To ensure that $\alpha_{\rm w}$ is not dominated by the brightest pixels, we also compute the unweighted analog $\alpha_{\rm uw}$ using Eq.~\ref{moment_for_alpha} without flux weighting, which instead captures the symmetry of the overall nebular morphology. 
Fig.~\ref{alpha_vs_sigma} also presents the $\alpha_{\rm uw}$ profile. 
The $\alpha_{\rm uw}$ profiles (Fig.~\ref{alpha_vs_sigma}) corroborate the trend: Ly$\alpha$ nebulae around obscured quasars are systematically more symmetric than those around unobscured quasars under the 1-$\sigma$ scatter. 

We obtain consistent results using a Fourier decomposition method to quantify nebular symmetry \citep{Brok2020,Wang2023}.
In this approach, the SB image is expressed in polar coordinates as ${\rm SB}(r,\theta)$, and a Fourier transform is applied (Eq.~\ref{fourier_SB}):
\begin{equation}
{\rm SB}(r,\theta)=\Sigma_{k=0}^{\infty}[a_{k}(r)cos(k\theta)+b_{k}(r)sin(k\theta)]
\label{fourier_SB}
\end{equation}
where the Fourier coefficients $a_{k}(r)$ and $b_{k}(r)$ are computed from the SB image via Eq.~\ref{fourier_coefficient}.
\begin{subequations}
\label{fourier_coefficient}
    \begin{align}
        &  a_{k}(r)=\frac{1}{2\pi}\int_{0}^{2\pi}SB(r,\theta)cos(k\theta)d\theta\\
        &  b_{k}(r)=\frac{1}{2\pi}\int_{0}^{2\pi}SB(r,\theta)sin(k\theta)d\theta
    \end{align}
\end{subequations}
The deviation from circular symmetry at a given radius is quantified by $c_{k}(r)$ (Eq.~\ref{fourier_ck_eq}): 
\begin{equation}
    c_{k}(r)=\sqrt{a_{k}(r)^{2}+b_{k}(r)^{2}}
    \label{fourier_ck_eq}
\end{equation}
The fractional amplitude $c_{k}(r)/c_{0}(r)$ measures the significance of the asymmetry at radius $r$. 
Following \citet{Brok2020}, we define $c_{\rm asymmetry}=(c_{1}/c_{0})^{2}+(c_{2}/c_{0})^{2}$, neglecting higher-order coefficients that are negligibly small compared to $c_{1}$ and $c_{2}$.
Larger values of $c_{\rm asymmetry}$ correspond to greater asymmetry.

To verify the robustness of these results, we further apply \textsc{CubExtractor} \citep{Cantalupo2019} to a purer subsample to re-subtract the PSFs and produce optimal-extraction (OE) images, on which we repeat the Fourier decomposition. 
This subsample is constructed by excluding quasars from programs (iii) and (v): program (v) is biased toward radio-loud quasars, while program (iii) targets star-forming galaxies (SFGs) that have been reclassified as AGNs based on the BPT diagram.
Quasars residing in overdense regions or with close companions are also excluded.

\begin{figure}
\centering
\includegraphics[width=\columnwidth]{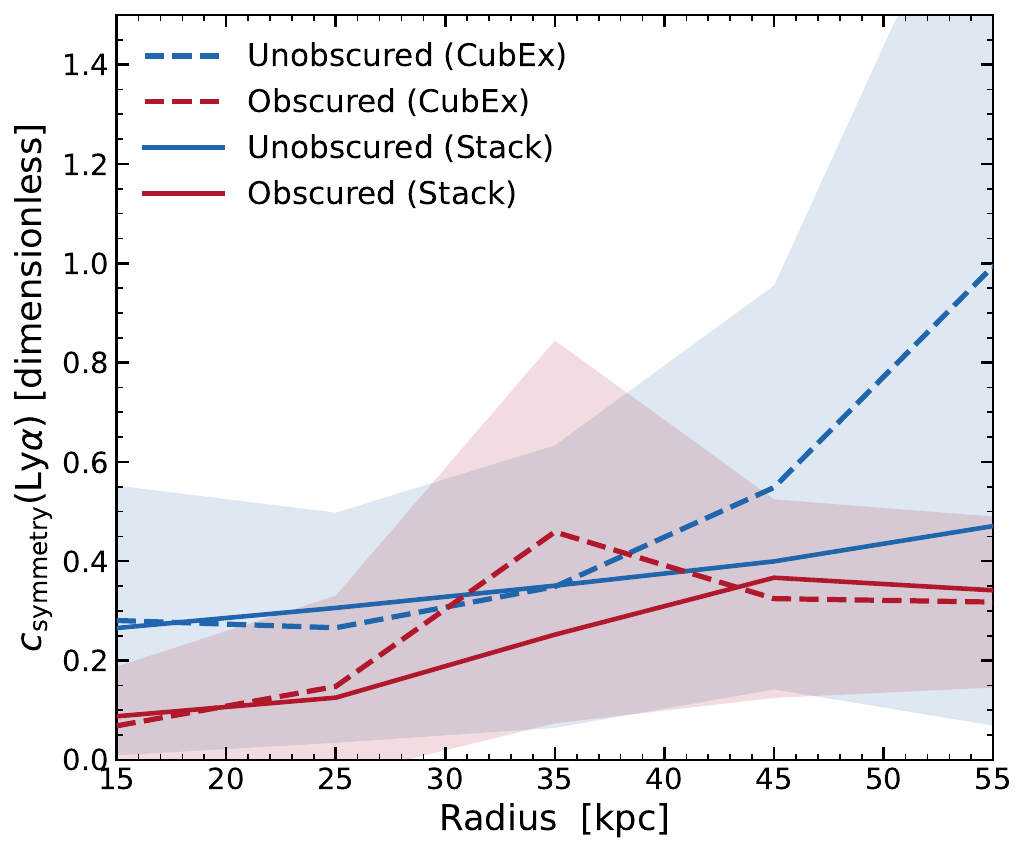}
\caption{Radial profiles of $c_{\rm symmetry}$, computed using Eqs.~\ref{fourier_coefficient} and \ref{fourier_ck_eq}. 
Shaded regions indicate the $1\sigma$ scatter. 
Higher $c_{\rm symmetry}$ values correspond to lower nebular symmetry. 
For each nebula, $c_{\rm symmetry}$ is computed from both the normally stacked (solid lines) and OE (dashed lines) images; the mean values are shown, with different colors denoting distinct quasar types. 
Both panels (a) and (b) reveal a consistent trend: Ly$\alpha$ nebulae around unobscured quasars are less symmetric than those around obscured quasars.}
\label{fourier_decomposition}
\end{figure}

The resulting radial profiles of $c_{\rm asymmetry}$ are shown in Fig.~\ref{fourier_decomposition}. 
Values derived from the standard stacked images and the OE images agree within the 1-$\sigma$ scatter over 15–55 kpc. 
Both approaches reveal the same trend: despite considerable scatter, nebulae around obscured quasars generally exhibit lower $c_{\rm asymmetry}$—and hence more symmetric morphologies—than those around unobscured quasars within $\sim50$ kpc, although the difference is modest.
This result contrasts with the unified scheme, which predicts that Ly$\alpha$ nebulae around unobscured quasars should be more symmetric, with $\alpha_{\rm w, unobs}=0.7-0.6$, compared to $\alpha_{\rm w, obs}=0.6-0.4$ for nebulae around obscured quasars (dashed line in Fig.~\ref{alpha_vs_sigma}). 

Taken together, the $\alpha_{\rm w}$, $\alpha_{\rm uw}$, and Fourier decomposition analyses all consistently indicate that Ly$\alpha$ nebulae around unobscured quasars are less symmetric than those around their obscured counterparts. 
This finding is at odds with the unified scheme, which expects Ly$\alpha$ nebulae around unobscured quasars to be more symmetric ($\alpha_{\rm w, unobs}=0.7-0.6$) than those around obscured quasars ($\alpha_{\rm w, obs}=0.6-0.4$) \citep{Zhang2025a}. 
We discuss this tension in detail in Sec.~\ref{sec:discussion}.

\subsection{Nebular surface brightness (SB) profile} \label{nebular_SB}
\begin{figure*}
\centering
\includegraphics[width=\textwidth]{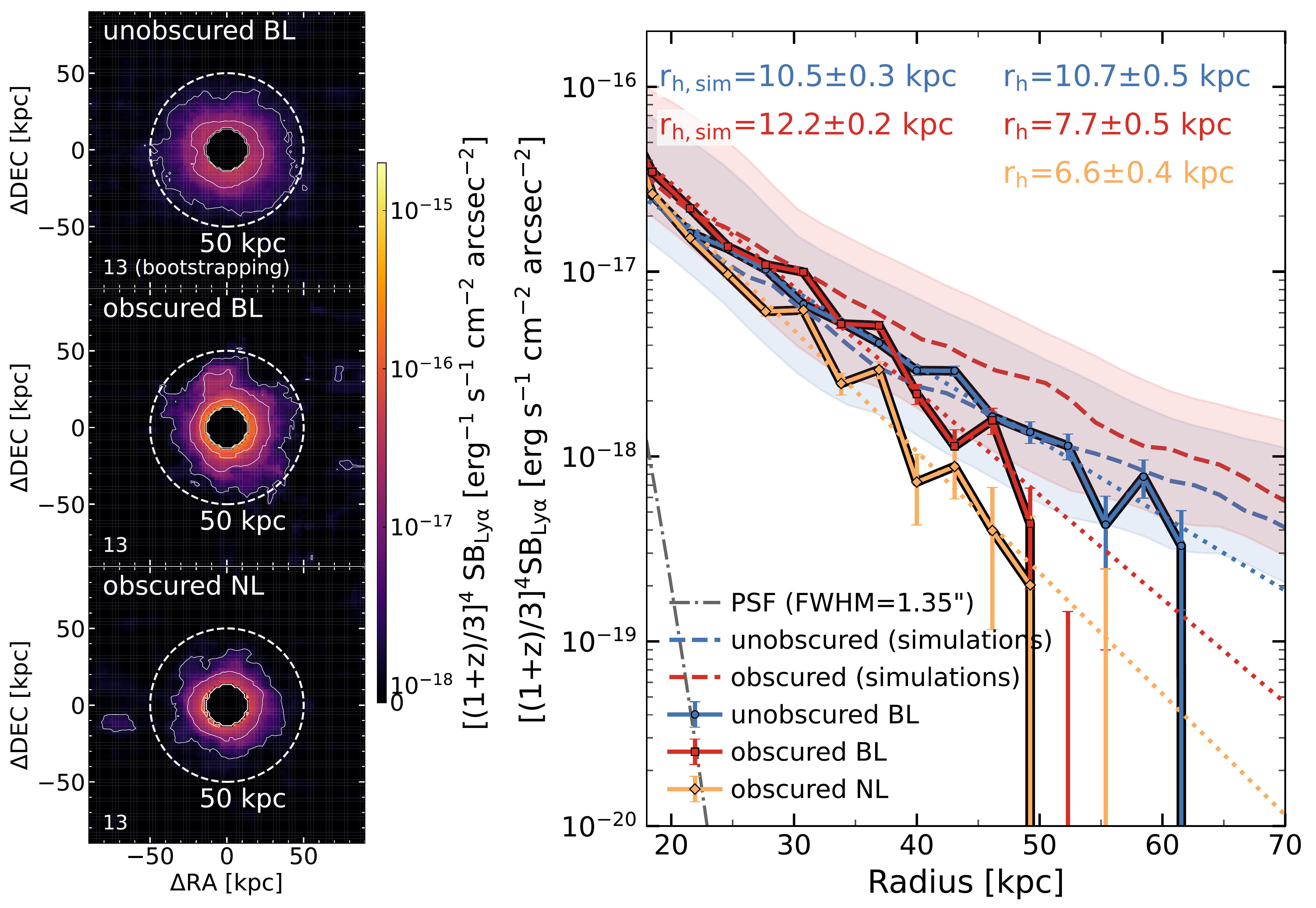}
\caption{The stacked images and SB profiles of Ly$\alpha$ nebulae around different types of quasars. 
{\bf Left:} The NB images are created from datacubes with a spectral width of $\pm 1500$ km s$^{-1}$. 
Images under the same quasar types are stacked.  
The white contours represent the SB noise levels of 2-$\sigma_{\rm SB}$, 10-$\sigma_{\rm SB}$, and 50-$\sigma_{\rm SB}$, respectively. 
The white dashed circle marks a radius of 50 kpc. 
Numbers at the bottom-left corner indicate the number of individual images used for stacking. 
We apply the bootstrapping technique to images of unobscured quasars with an equal number to the obscured quasars.
{\bf Right:} The SB profiles of Ly$\alpha$ nebulae around different types of quasars. 
Different colors denote different quasar types. 
The errorbar represents the 1-$\sigma$ uncertainty obtained via error propagation.
The dotted lines denote the best fit with an exponential function. 
The black dash-dotted line denotes the point spread function (PSF) with FWHM=$1.35''$ which is normalized to the peak flux of the obscured BL quasars. 
The dashed lines represent the simulated SB profiles under the standard AGN unified model with the shaded regions denoting a range of 16th-84th. 
The cosmic dimming effect is corrected for both the stacked images and profiles. 
The images show that Ly$\alpha$ nebulae around unobscured BL quasars extend to $\approx 50$ kpc, while nebulae around obscured quasars extend less than this radius. 
This is because nebulae around obscured quasars have steeper profiles ($r_{\rm h}\approx 7.2$ kpc) compared to nebulae around unobscured quasars ($r_{\rm h}= 10.7$ kpc). 
This contrasts with simulations based on the AGN unified model, which predict that nebulae around unobscured quasars should be more compact ($r_{\rm h,sim}$=10.5 kpc) than those around obscured quasars ($r_{\rm h,sim}$=12.2 kpc). }
\label{nebular_SBprofile}
\end{figure*}

We apply the median stacking technique to construct composite images and SB profiles.
To ensure a fair comparison, we use a bootstrapping approach to randomly draw a subsample of unobscured quasars equal in number to the obscured quasar sample.
All images have been corrected for the cosmic dimming effect to a reference redshift of $z=2$. 
As shown in Fig.~\ref{nebular_SBprofile}, nebulae around obscured quasars appear less extended than those around unobscured quasars.
This difference is driven by a steeper decline in nebular brightness around obscured quasars, as evidenced by their SB profiles. 
We quantify this behavior by fitting the profiles with an exponential function of the form $\rm SB(r)=Ce^{-r/r_{\rm h}}$, where $C$ is the normalization and $r_{\rm h}$ is the scale length (i.e., the slope). 

Unobscured quasars yield a larger scale length ($r_{\rm h}=10.7\pm 0.5$ kpc) than their obscured counterparts ($r_{\rm h, NL}=6.6\pm 0.4$ kpc and $r_{\rm h, BL}=7.7\pm 0.5$ kpc), with the difference exceeding the $1\sigma$ fitting uncertainties. 
Interestingly, simulations based on the standard AGN unified model predict the opposite trend: if the intrinsic quasar properties are consistent, nebulae around unobscured quasars should exhibit steeper profiles ($r_{\rm h}=10.5\pm0.3$ kpc) than those around obscured quasars ($r_{\rm h}=12.2\pm0.2$ kpc) due to projection effects arising from the AGN ionization cone \citep{Zhang2025a}. 
We return to this tension between the observations and the unified-model predictions in Sec.~\ref{sec:discussion}.

\subsection{Nebular velocity dispersion profile} \label{nebular_velocity_dispersion}
\begin{figure*}
\centering
\includegraphics[width=\textwidth]{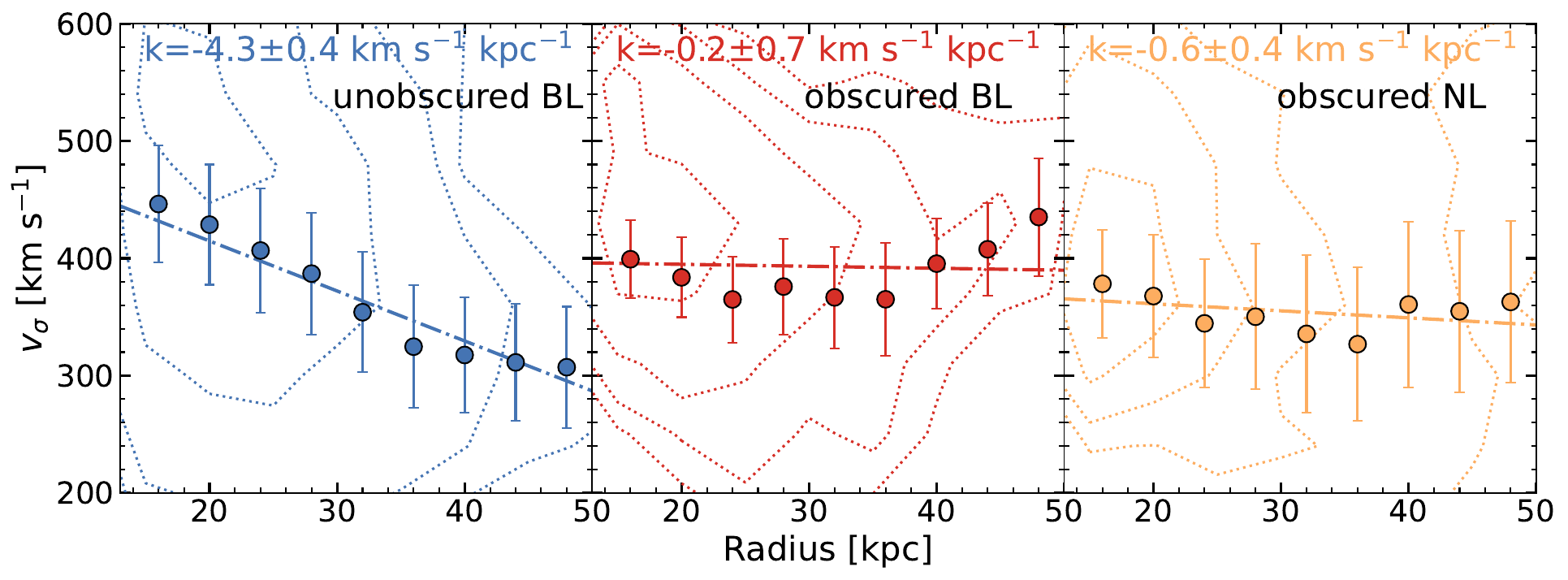}
\caption{{\bf Left:} The radial profile of the flux-weighted velocity dispersion of Ly$\alpha$ nebulae around unobscured BL quasars. 
Only pixels with $S/N\geq 5$ are selected. 
The dotted contours denote the 2-dimensional (2D) histogram of these pixels. 
The filled dots denote the median value in each radial bin. 
The errorbar represents the 1-$\sigma$ scatter.  
A dot-dashed line shows the best linear fit to these points with the slope shown at the top. 
{\bf Middle \& Right:} Same as the left panel, but for obscured BL and NL quasars, respectively. 
We find that nebulae around unobscured quasars have strongly declining dispersion profile in contrast to the flat profiles of nebulae around obscured quasars. }
\label{nebular_vdisp}
\end{figure*}

Beyond the SB and morphological asymmetry presented above, we also examine the kinematics of the Ly$\alpha$ nebulae through their velocity dispersion profiles.
For each datacube, we construct a flux-weighted velocity dispersion map following the approach of \citet{Zhang2023a}, retaining only high-significance pixels with $S/N\geq 5$. 
The resulting radial profiles are shown in Fig.~\ref{nebular_vdisp}. 

Nebulae around unobscured BL quasars exhibit a pronounced decline in velocity dispersion with radius, characterized by a slope of $-4.3\pm 0.4$ km s$^{-1}$ kpc$^{-1}$. 
The tightness of this trend is confirmed by a Spearman correlation coefficient of $R_{s}=-0.98$ with $p\approx 0\%$. 
In contrast, nebulae around obscured quasars—both BL and NL—display remarkably flat dispersion profiles, with shallow slopes of $-0.2\pm 0.7$ km s$^{-1}$ kpc$^{-1}$ and $-0.6\pm 0.4$ km s$^{-1}$ kpc$^{-1}$, respectively, indicating no significant radial gradient. 
We verify that the declining trend for unobscured quasars is not an artifact of the larger sample size by applying bootstrapping to randomly stack only 13 profiles (matching the number of obscured BL quasars).
Even with this reduced sample, the trend remains robust, yielding a slope of $\approx -3.4\pm 0.7$ km s$^{-1}$ kpc$^{-1}$ and $R_{s}=-0.79$ ($p\approx 0.6\%$).

Hydrodynamical simulations associate such declining dispersion profiles with the presence of large-scale outflows \citep{Fielding2020}, which—in the framework of the unified model—should be present around both quasar types. 
The unambiguous declining profile observed in unobscured quasars, which is consistent with recent measurements of Ly$\alpha$ nebulae at $z\approx 3$ \citep{Lobos2025}, stands in stark contrast to the flat profiles of their obscured counterparts.
This dichotomy indicates that large-scale outflows are prevalent exclusively in the unobscured population, implying fundamentally different circumgalactic gas environments for the two quasar types.

\section{Discussion} \label{sec:discussion}
In this section, we explore the connections between nebulae and AGN properties in Sec.~\ref{nebulae_AGN_connection}. 
Then we discuss whether such connections can be reproduced by the unified model in Sec.~\ref{unified_model_comparison} and and evolutionary scenario in Sec.~\ref{evolutionary_scenario}. 
Finally, we discuss what we can gain from the line ratio diagnostics in Sec.~\ref{line_ratio}. 

\subsection{Connection between nebular and AGN properties} \label{nebulae_AGN_connection}
The results presented in Sections~\ref{nebular_morphology}—\ref{nebular_velocity_dispersion} reveal that Ly$\alpha$ nebulae around unobscured and obscured quasars differ markedly in morphology, surface brightness, and kinematics. 
A critical question is whether these differences are intrinsically driven by the level of quasar obscuration—which would constitute a departure from the predictions of the unified scheme—or whether they instead arise indirectly through correlations with secondary AGN properties such as the SMBH mass ($M_{\rm BH}$) or radio loudness ($R_{K}$). 

\begin{figure*}
\centering
\includegraphics[width=\textwidth]{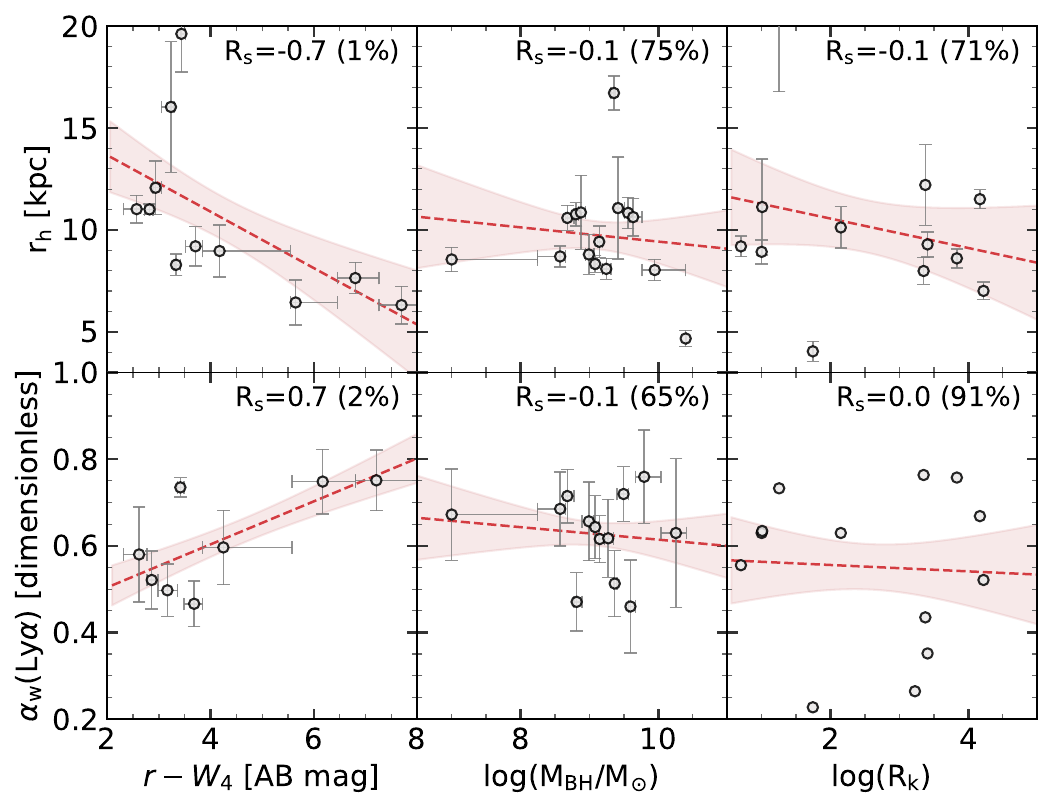}
\caption{Correlation between Ly$\alpha$ nebular parameters and quasar properties. 
{\bf Upper left} $r_{\rm h}$ as a function of $r - W4$, a proxy for quasar obscuration. 
Horizontal and vertical error bars represent the bin size and 1-$\sigma$ scatter, respectively. 
The red dashed line indicates the best linear fit, with the shaded area showing the 1-$\sigma$ fitting uncertainty. 
The Spearman correlation coefficient and the corresponding p-value are displayed in the top-right corner. 
{\bf Upper middle \& right:} same as upper left but showing the $r_{\rm h}$ as a function of SMBH mass and radio loudness, respectively. 
{\bf Bottom:} mirror those in the upper panels but presenting the dependence of nebular asymmetry ($\alpha_{\rm w}$) on quasar properties. 
These results indicate that significant correlations—defined by p-values $\leq 5\%$—exist only between the nebular parameters and the quasar obscuration.}
\label{nebulae_AGN_connections}
\end{figure*}

We analyze the variation in nebulae properties in relation to obscuration, $M_{\rm BH}$, and $R_{K}$. 
Obscuration is quantified by $r-W_4$, while $M_{\rm BH}$ and $R_{K}$ are derived as described in Sec.~\ref{data_sample_phys_properties}. 
We bin the sample by $r-W_{4}$, $M_{\rm BH}$, and $R_{\rm K}$, computing the median $\alpha_{\rm w}$ and fitting the median-stacked SB profile to derive $r_{\rm h}$ in each bin. 
Connections between nebular and AGN properties are summarized in Fig.~\ref{nebulae_AGN_connections}.  

Both the Spearman correlation coefficients ($R_{\rm s}$) and the linear fits yield consistent conclusions: nebular properties exhibit a strong dependence on $r-W_{4}$ ($|R_{\rm s}|\approx 0.7$, $p\leq 2.5\%$), but show no significant dependence on $M_{\rm BH}$ or $R_{K}$ ($|R_{\rm s}|\leq 0.2$, $p\geq 62\%$). 
These results indicate that the extent and symmetry of Ly$\alpha$ nebulae are primarily governed by the degree of quasar obscuration, while SMBH mass and radio loudness play at most a minor role.
This finding is broadly consistent with earlier studies of Ly$\alpha$ nebulae around unobscured quasars at $z\approx 3$, which likewise reported no significant correlation between nebular SB profile slope or symmetry and either quasar luminosity or radio loudness \citep{Arrigoni2019,Mackenzie2021}.

To verify that these conclusions are robust against the choice of binning, we further examine the stacked SB profiles and $\alpha_{\rm w}$ values across different subdivisions of $r-W_{4}$, $M_{\rm BH}$, and $R_{K}$ (restricted to unobscured quasars for the latter two). 
We also compare the nebular properties of red quasars with and without [\oiii] detections to assess whether potential outflows introduce additional heterogeneity into the obscured sample. 

\begin{figure*}
\centering
\includegraphics[width=\textwidth]{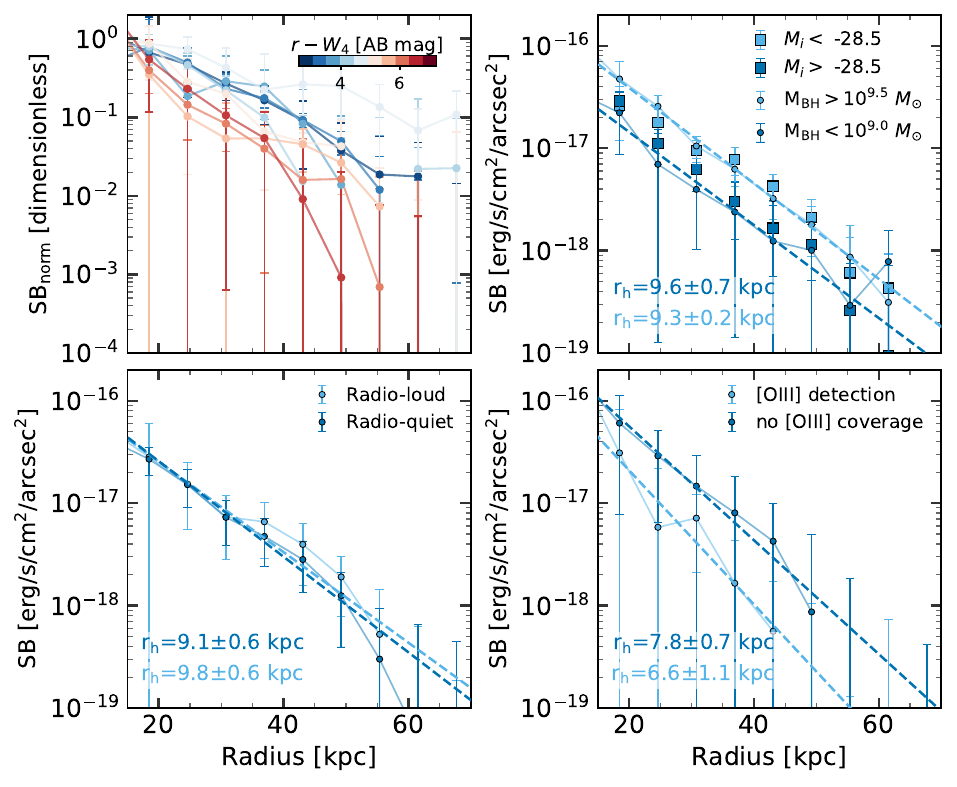}
\caption{The stacked SB profile obtained by dividing our sample into different bins.
{\bf Upper left:} The stacked SB profiles in different bins of $r-W_{4}$ which is denoted by the color. 
The SB is normalized by the flux at $r=15$ kpc. 
The errorbar represents the 1-$\sigma$ scatter. 
{\bf Upper right \& bottom left:} Same as panel (a) but for the stacked profile in different bins of SMBH mass, absolute $i-$band magnitude, and radio loudness. 
{\bf Bottom right:} Stacked profiles of nebulae from program (v) in Sec.~\ref{sec:selection}. 
The quasars are divided into two bins: quasars with [\oiii] detections and those without [\oiii] coverage. 
These panels show that $r_{\rm h}$ correlates significantly only with $r-W_{4}$, and not with $M_{\rm BH}$, $M_{i}$, or radio loudness, consistent with the results shown in Fig.~\ref{nebulae_AGN_connections}. 
Panel (d) indicates that red quasars with [\oiii] detections have similar $r_{\rm h}$ to those without [\oiii] coverage.}
\label{rebin_sb_profile}
\end{figure*}

The stacked and normalized SB profiles are presented in Fig.~\ref{rebin_sb_profile}. 
The upper-left panel reveals a clear and continuous trend: as $r-W_{4}$ increases—that is, as the quasar becomes more obscured—the SB profile steepens, corroborating the conclusion that more obscured quasars are surrounded by more compact nebulae. 
Turning to $M_{\rm BH}$, we split the unobscured sample into two bins ($M_{\rm BH}>10^{9.5}\ M_{\odot}$ and $M_{\rm BH}<10^{9.0}\ M_{\odot}$). 
While the resulting profiles share nearly identical scale lengths, the higher-mass bin exhibits a systematically elevated SB, implying that more massive—and hence more luminous—quasars illuminate brighter nebulae without significantly altering their spatial extent. 

Because $M_{\rm BH}$ is derived from the SED-fitted bolometric luminosity and is therefore subject to considerable uncertainty, we cross-check this result using the absolute $i$-band magnitude $M_{i}$ as an independent luminosity proxy (squares in panel~b of Fig.~\ref{rebin_sb_profile}). 
Binning by $M_{i}$ reproduces the same pattern, consistent with previous observations of Ly$\alpha$ nebulae at $z\approx 3$ \citep{Mackenzie2021,Lobos2025} and lending further support to the reliability of the SED-derived $M_{\rm BH}$.

Regarding radio loudness, we find no significant difference between the radio-loud and radio-quiet bins.
The sole distinction is a marginally larger $r_{\rm h}$ in the radio-loud bin, with a residual of $\Delta r_{\rm h}\approx 1.2$ kpc—far too small to account for the observed disparity between nebulae around obscured and unobscured quasars (Fig.~\ref{rebin_sb_profile}).
This offset is nevertheless broadly consistent with earlier reports of slightly more extended nebulae around radio-loud unobscured quasars \citep{Arrigoni2019}. 
For the red quasars, the bin with [\oiii] detections shows a marginally less extended profile, which can be attributed to their higher obscuration ($r-W_{4}=7.7\pm0.5$) compared to those without [\oiii] coverage ($r-W_{4}=7.0\pm0.6$).

\begin{figure}
\centering
\includegraphics[width=\columnwidth]{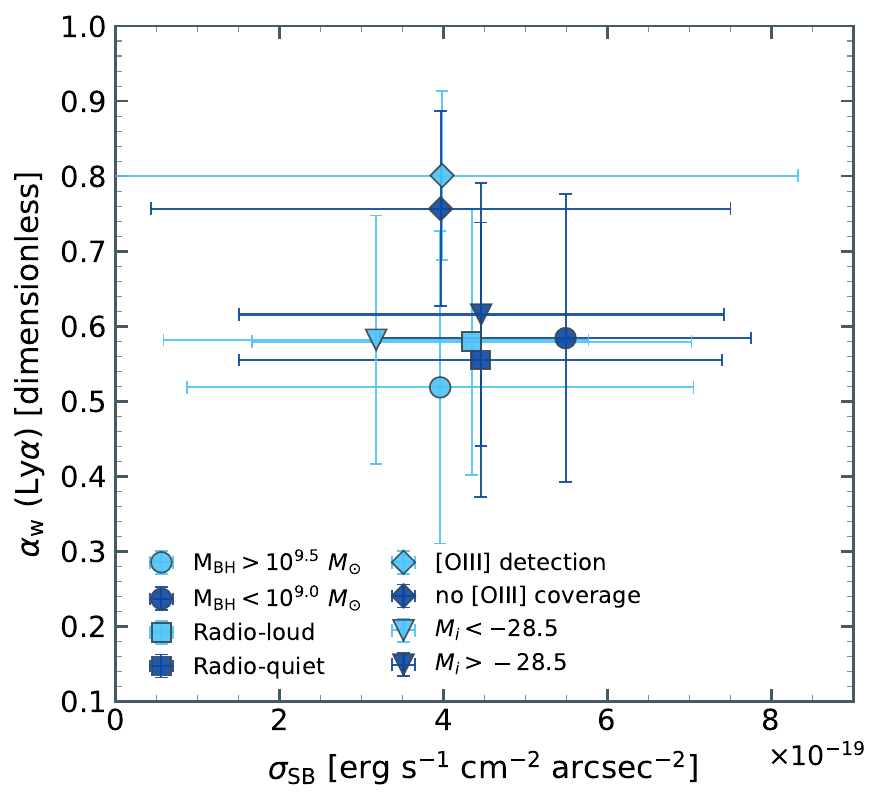}
\caption{The $\alpha_{\rm w}$ obtained by dividing our sample into different bins. 
The sample is binned in the same way as Fig.~\ref{rebin_sb_profile}, with different markers representing different bins. 
The errorbar denotes the 1-$\sigma$ scatter. 
This figure demonstrates that the $\alpha_{\rm w}$ remains consistent across bins within 1-$\sigma$, suggesting no correlation with SMBH mass, absolute $i$-band magnitude, or radio loudness. 
Besides, nebulae around red quasars with [\oiii] detections show similar $\alpha_{\rm w}$ values to those without [\oiii] coverage, suggesting that including nebulae with [\oiii] detections in $\alpha_{\rm w}$ calculation is justified.}
\label{rebin_nebular_morphology}
\end{figure}

Adopting the same binning scheme, we also compute the median $\alpha_{\rm w}$ in each bin (Fig.~\ref{rebin_nebular_morphology}).
The nebular symmetry remains consistent within the 1-$\sigma$ scatter across all bins of $M_{\rm BH}$, $R_{K}$, and [\oiii] detection.
No clear trend with absolute $i$-band magnitude is observed either (triangles in Fig.~\ref{rebin_nebular_morphology}).
Collectively, this analysis confirms that neither $r_{\rm h}$ nor $\alpha_{\rm w}$ exhibits a significant dependence on $M_{\rm BH}$ or radio loudness.
It further demonstrates that red quasars with and without [\oiii] detections share broadly consistent nebular profiles and symmetry, justifying their combined inclusion in the obscured sample. 

Because Ly$\alpha$ is a resonant line, its spatial distribution may not directly trace the underlying gas \citep{Costa2022,Zhang2025a}. 
Assessing the role of resonant scattering is therefore essential.
Simulations show that resonant scattering can increase both $r_{\rm h}$ and $\alpha_{\rm w}$ by isotropically redistributing Ly$\alpha$ photons outward \citep{Zhang2025a,Michel2020}. 
If resonant scattering dominated, these quantities would vary non-monotonically with obscuration: increasing obscuration would raise the neutral hydrogen fraction and enhance scattering (producing more extended, isotropic nebulae), while in the heavily obscured regime the limited ionizing budget would cause nebulae to contract.

Instead, we observe a clear monotonic relation between nebular parameters and $r-W_{4}$ (Fig.~\ref{rebin_nebular_morphology}), indicating that resonant scattering has limited influence and that the correlations are more plausibly driven by the actual CGM gas distribution linked to quasar obscuration.


\subsection{Compare with the AGN unified model predictions} \label{unified_model_comparison}
As demonstrated in Sec.~\ref{nebular_morphology} and Sec.~\ref{nebular_SB},
our observations reveal that Ly$\alpha$ nebulae around obscured quasars are
systematically more symmetric and more compact than those around unobscured
quasars—the opposite of what is predicted by the unified-model simulations of
\citet{Zhang2025a}.  
Those simulations assume that obscured and unobscured quasars share identical intrinsic properties, specifically $M_{\rm BH}$ and the dusty-torus opening angle. 
Although the SED fitting in Sec.~\ref{data_sample_phys_properties} broadly supports these assumptions, the SED-derived parameters carry substantial uncertainties.  
To account for this, we relax the assumptions and perform a bootstrap resampling of the mock observations, allowing both $M_{\rm BH}$ and the torus opening angle to vary within the ranges spanned by the simulations.

\begin{figure}
\centering
\includegraphics[width=\columnwidth]{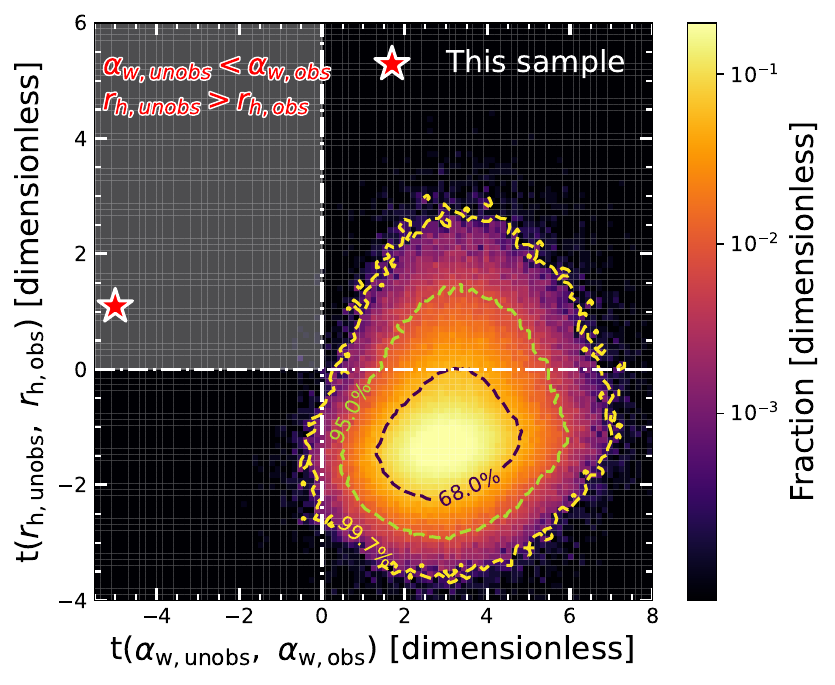}
\caption{The 2D histogram of t-statistics comparing the $r_{\rm h}$ and $\alpha_{\rm w}$ of different types of quasars. 
We generate 500000 bootstrap samples of 59 and 26 mock nebulae around unobscured and obscured quasars, respectively, based on the AGN unified model. 
For each sample, we compute $r_{\rm h}$ and $\alpha_{\rm w}$ and apply the Welch t-test. 
The contours indicate the enclosed fractional levels. 
$\approx 90\%$ of iterations yield $t(\alpha_{\rm w, unobs},\alpha_{\rm w, obs})\geq 0$ and  $t(r_{\rm h, unobs},r_{\rm h, obs})\leq 0$, which is opposite to the observed results (red star). 
Only $3.2\times 10^{-3}\%$ iterations are consistent with our findings. 
This indicates that the standard AGN unified model is challenging in reproducing the observed results.}
\label{unified_predictions}
\end{figure}

Starting from the 180 mock nebulae (90 per quasar type) generated by
\citet{Zhang2025a}, we draw 500000 bootstrap samples with replacement. 
Each iteration randomly selects 59 unobscured and 26 obscured mock nebulae,
matching the sample sizes of our observed dataset. 
For every resampled set we compute $\alpha_{\rm w}$—using the same SB threshold of
$8.0\times 10^{-19}$ erg s$^{-1}$ cm$^{-2}$ arcsec$^{-2}$ adopted for the
observations—and $r_{\rm h}$, and evaluate the significance of the
differences between the two populations with the Welch $t$-test. 
Positive $t$-values indicate that unobscured quasars possess larger
$\alpha_{\rm w}$ and $r_{\rm h}$ than their obscured counterparts.
The resulting 2D distribution of $t$-statistics is presented in
Fig.~\ref{unified_predictions}.

The bootstrap analysis yields a $\approx 90\%$ probability that, under the unified model, nebulae around unobscured quasars are simultaneously more symmetric and more extended than those around obscured quasars—a trend that is diametrically opposed to our measurements.  
Only $3.2\times 10^{-3}\%$ of the iterations reproduce the observed sign combination.
Although $r_{\rm h,unobs}\geq r_{\rm h,obs}$ occurs in $\sim 10\%$ of iterations, the vast majority of these cases have $\Delta r_{\rm h}\leq 1$ kpc, well below the observed offset of $\Delta r_{\rm h}\geq 2$ kpc (Fig.~\ref{nebular_SBprofile}); 
merely $0.22\%$ of iterations reach $\Delta r_{\rm h}\geq 1$ kpc. 
Taken together, these statistics demonstrate that the standard AGN unified model has great difficulty in accommodating our observational findings. 

We note that the mock observations are drawn from FIRE-1 cosmological simulations, which do not include AGN feedback \citep{Hopkins2014,Feldmann2016,Angles2017,Feldmann2017}.  
Assessing how feedback might alter the bootstrap results is therefore necessary. 
Both radiation-driven and jet-driven feedback operate preferentially along the
polar axis of the torus \citep{Schmitt2002,Drouart2012,Burillo2021,Ruffa2020}, introducing anisotropy in the thermodynamic state and ionization level of the CGM. 
On the one hand, simulations show that feedback can raise gas temperatures and
lower densities by $0.1-0.3$ dex \citep{Terrazas2020,Truong2021,Yang2024}, suppressing Ly$\alpha$ emissivity by up to 0.4 dex along the polar direction \citep{Zhang2025a}. 
On the other hand, the same feedback can substantially ionize neutral hydrogen \citep{Byrohl2021,Costa2022,Meenakshi2022}, potentially boosting Ly$\alpha$ emission by a factor of $\approx 10$ \citep{Costa2022}. 
Beyond modifying the gas thermodynamics, AGN-driven outflows can shock-excite CGM gas and transport dense ISM material to distances of several tens of kpc \citep{Li2021,Venturi2023,Zhang2023a,Villar2024}, further enhancing Ly$\alpha$ emissivity.

Radiation-driven feedback, in particular, has been shown to increase—rather than diminish—the overall nebular brightness \citep{Byrohl2021}. 
Because this enhancement is stronger within the ionization cone than outside it, radiation-driven feedback would likely amplify the predicted differences between the two quasar populations under the unified framework. 
The net effect of jet-driven feedback is harder to assess, since current nebular simulations do not incorporate it. 
However, because its impact on thermodynamic properties and ionization state push emissivity in opposite directions, any change in the nebular contrast between the two populations is probably modest. 
Overall, the considerations above suggest that including AGN feedback would enhance—or at least not significantly reduce—the nebular differences predicted by the unified model.

We caution that the influence of different AGN feedback mechanisms warrants more detailed investigation. 
Moreover, the scattering of Ly$\alpha$ photons from the broad-line region is treated in the \textsc{cloudy} calculations only approximately, via continuum pumping \citep{Ferland2017}. 
More sophisticated simulations are therefore needed to draw firmer conclusions.

\subsection{The evolutionary scenario} \label{evolutionary_scenario}
\begin{figure*}
\centering
\includegraphics[width=\textwidth]{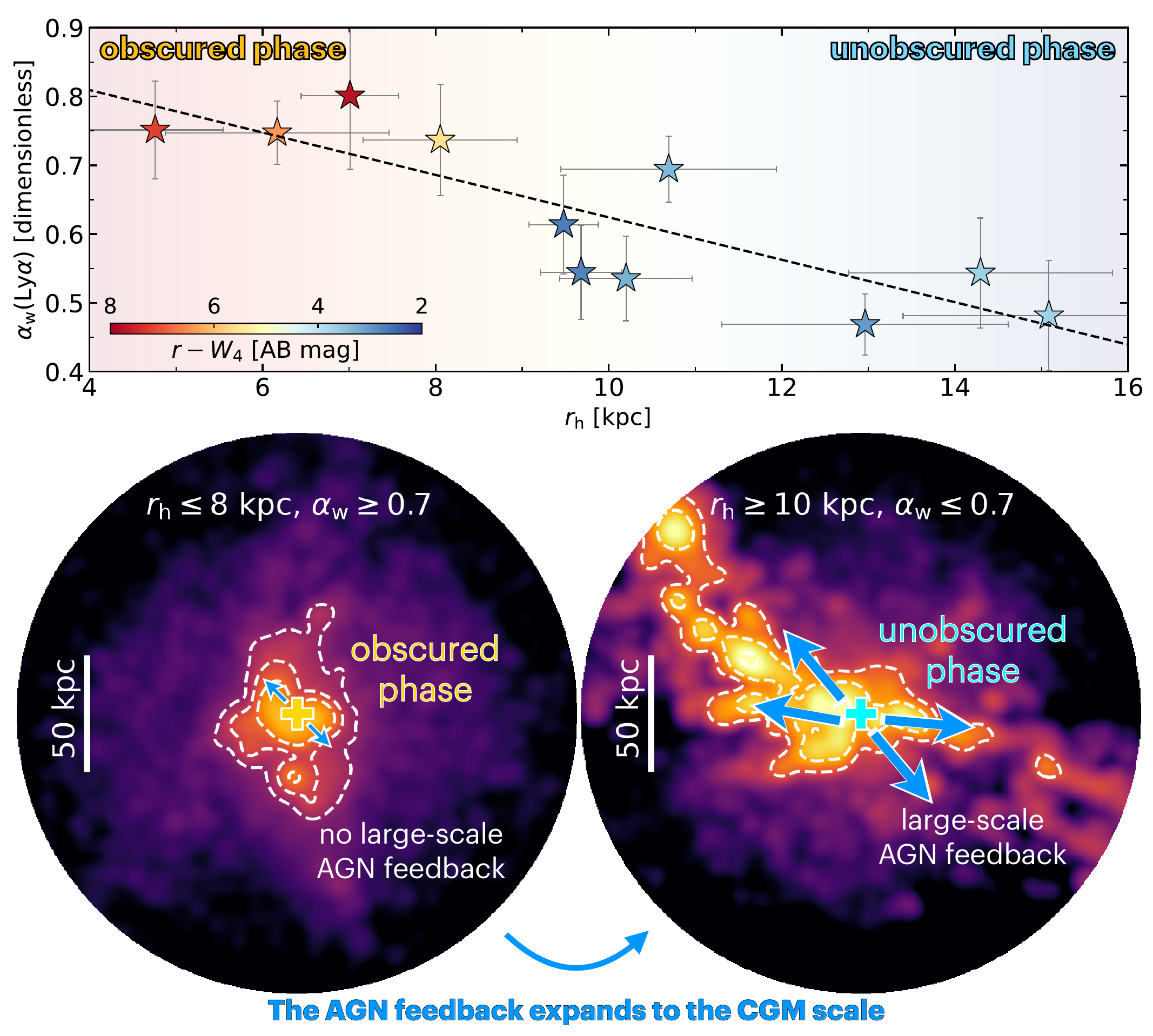}
\caption{Illustration of the Ly$\alpha$ nebulae under the evolutionary scenario. 
{\bf Upper:} the $\alpha_{\rm w}$ as a function of $r_{\rm h}$ with the color indicating the $r-W_{4}$. 
The errorbar denotes the 1-$\sigma$ scatter. 
The dashed line is the linear fit. 
{\bf Bottom:} A schematic illustrating the evolution of Ly$\alpha$ nebulae as the quasar transitions from the obscured to the unobscured phase under the evolutionary scenario. 
Blue arrows depict the AGN feedback, while white contours outline the nebulae. 
As the AGN feedback expands from the galactic to CGM scale, it not only drives the quasar’s transition from a red, faint, obscured phase to a blue, bright, unobscured phase by expelling dust from the ISM or circumnuclear regions—but also reshapes the CGM gas from a compact and symmetric structure to a more diffuse and asymmetric one. 
This scenario naturally results in a decrease in nebular symmetry ($\alpha_{\rm w}$ drops from 0.8 to 0.5) and an increase in nebular extent ($r_{\rm h}$ rises from 6 kpc to 14 kpc) as the quasar becomes more unobscured ($r-W_{4}$ drops from 7 to 3 AB magnitudes), as shown in the upper panel.}
\label{evolutionary_scenario_picture}
\end{figure*}

A growing body of observational evidence calls into question whether the standard AGN unified model—purely orientation-based—can adequately describe quasars at $z\geq 2$ \citep{Harikane2023,Matthee2024,Maiolino2024}.
An alternative framework, the evolutionary scenario, posits that AGN feedback progressively expels dust and gas from the nuclear environment, driving a transition from an obscured to an unobscured phase \citep{Sanders1988,Hopkins2008,Alexander2012}.
In this picture, unobscured quasars represent a later evolutionary stage in which feedback has had more time to propagate outward, whereas obscured quasars correspond to an earlier phase in which the feedback has yet to reach large galactic scales (Fig.~\ref{evolutionary_scenario_picture}). 
This framework naturally accounts forthe observed differences reported in Sec.~\ref{sec:results}.

First, the more extended SB profiles measured around unobscured quasars ($r_{\rm h}=10.7\pm0.5$ kpc, Sec.~\ref{nebular_SB}) are consistent with simulations showing that AGN feedback can transport dense ISM and inner-CGM material out to tens of kiloparsecs \citep{Cielo2018,Costa2018,Barnes2020,Prasad2022}, effectively pushing gas to larger radii and flattening the SB profile during the later, unobscured phase.
This interpretation is supported by observations of large-scale AGN-driven outflows extending to $\sim100$ kpc around unobscured quasars at $z\geq 2$ \citep{Nesvadba2017,Travascio2020,Li2021}, as well as by [\oiii] imaging revealing that outflows around unobscured quasars tend to be more spatially extended than those around their obscured counterparts \citep{Perrotta2019,Tozzi2024,Lau2024}.

Second, the lower morphological symmetry of nebulae around unobscured quasar ($\alpha_{\rm w}=0.2-0.6$, Sec.~\ref{nebular_morphology})—compared to the more symmetric nebulae around obscured quasars ($\alpha_{\rm w}=0.6-0.8$)—is naturally explained by the anisotropic propagation of AGN feedback \citep{Nobels2022,Cresci2023,Yang2024}.
As the feedback expands into the CGM during the unobscured phase, it preferentially channels energy along certain directions, introducing additional anisotropy into the large-scale gas distribution.
In the earlier obscured phase, however, the feedback has not yet reached CGM scales, leaving the gas in a relatively undisturbed and hence more symmetric configuration.

Third, the kinematic signatures further support this picture. 
The declining trend in the unobscured population is consistent with feedback-driven turbulence that is strongest near the galaxy and weakens at larger radii \citep{Fielding2020}.
The flat profiles seen around obscured quasars, in contrast, indicate that large-scale feedback has not yet perturbed the CGM, consistent with their earlier evolutionary stage.

Taken together, these results suggest that the evolutionary scenario could be the more prevalent picture for the high-redshift quasars where the evolution of the CGM and AGN is tightly connected by the large-scale feedback. 
We note that, under this evolutionary picture, the obscuring material need not be confined to the circumnuclear region. 
Simulations indicate that dust and gas in the ISM can contribute hydrogen column densities and levels of obscuration comparable to those produced by circumnuclear tori \citep{Trebitsch2019}.
Recent observations reinforce this view, showing that ISM column densities can reach $\geq 10^{22}$ cm$^{-2}$ \citep{Circosta2019,Gilli2022,Andonie2024}. 

Crucially, while the AGN unified model is well-established in the local Universe, mounting evidence favouring the evolutionary scenario at $z\geq 2$ suggests a transition in the dominant AGN paradigm over cosmic time.
Given that $z=2-3$ coincides with the peak epoch of cosmic star formation \citep{Madau2014}, enhanced gas accretion onto galaxies could naturally lead to higher obscuration fractions than observed at low redshift.
This shift may be fundamentally linked to how massive halos hosting AGN accrete gas from the intergalactic medium.
Determining whether such a transition exists—and, if so, characterizing its nature—could provide critical insights into SMBH evolution and its role in shaping galaxies over cosmic history.

\subsection{Line Ratio diagnostic} \label{line_ratio}

The evolutionary scenario should also predict that the CGM surrounding unobscured quasars exhibit higher metallicity than that around obscured quasars, due to the enrichment from the large-scale AGN feedback \citep{Suresh2015}. 
Indeed, the CGM of unobscured quasars at $z\approx 3$ has been found to exhibit metallicities of $\approx 0.5 \ Z_{\odot}$ \citep{Guo2020}, whereas the metallicity of the CGM around obscured quasars remains poorly constrained.
Since a subset of our datacubes cover the \civ \ emission line, we can use the \civ/Ly$\alpha$ ratio as a metallicity proxy to test this prediction.

Among the datacubes with \civ \ coverage (18 unobscured and 11 obscured quasars), we detect \civ \ nebulae around 15 unobscured and seven obscured quasars.
For these detections, we measure the Ly$\alpha$ and \civ \ luminosities within an annulus of $15-30$ kpc and compute their ratio.
The inner radius is chosen to minimize PSF contamination, while the outer radius corresponds to the maximum extent of the detected \civ \ nebulae.
For quasars without \civ \ detections, we derive 3-$\sigma$ upper limits on the \civ \ luminosity within the same aperture and use these to place upper limits on the line ratio. 

\begin{figure*}
\centering
\includegraphics[width=\textwidth]{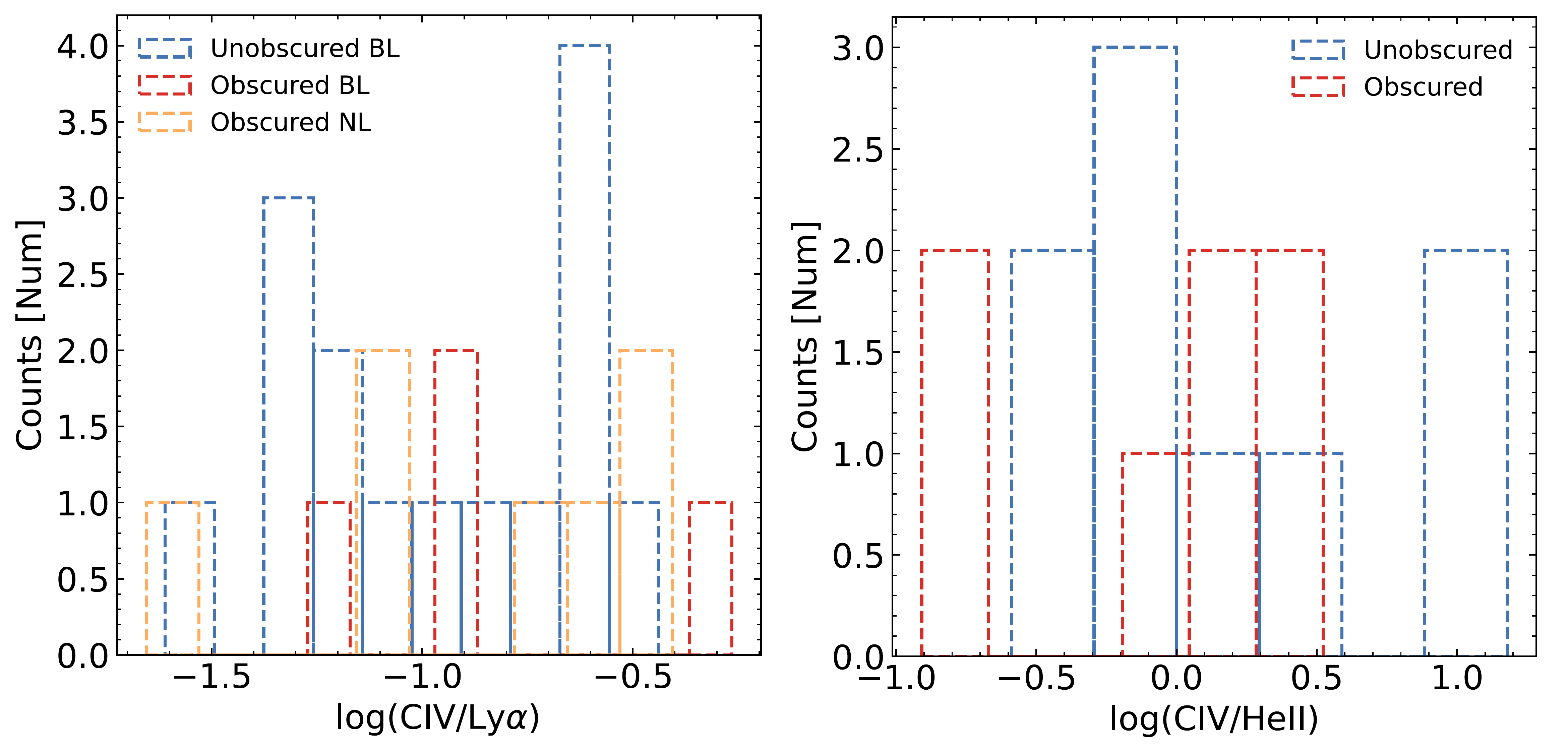}
\caption{The \civ/Ly$\alpha$ line ratio. {\bf Left:} 
The histogram of log(\civ/Ly$\alpha$) within $15-30$ kpc annulus, with colors indicating quasar types. 
The upper limits are also included. 
{\bf Right:} Same as the left panel, but for the line ratio of \civ/\heii. 
The blue and red histograms denote the unobscured and obscured quasars, respectively. 
Since these line ratios trace the gas metallicity, the results suggest similar CGM metallicity around different types of quasars. 
However, due to the resonant nature of Ly$\alpha$ and \civ \ lines, test with non-resonant lines like H$\alpha$, \heii \ and \ciii] is needed.}
\label{line_ratio_diagnostics}
\end{figure*}

The resulting distributions, including upper limits, are shown in Fig.~\ref{line_ratio_diagnostics}.
We measure median \civ/Ly$\alpha$ ratios of [\civ/Ly$\alpha]_{\rm unobs}=0.11\pm0.10$ and [\civ/\rm Ly$\alpha]_{\rm obs}=0.13\pm0.16$ for the unobscured and obscured populations, respectively.
Since this line ratio serves as a proxy for gas metallicity, the similarity between the two values would indicate no significant metallicity difference in the CGM surrounding the two populations.
However, both Ly$\alpha$ and \civ \ are resonant lines, making \civ/Ly$\alpha$ an imperfect metallicity diagnostic.
We therefore repeat the analysis using \heii—a non-resonant line—in place of Ly$\alpha$, deriving [\civ/\heii]$_{\rm unobs}=0.81\pm4.9$ and [\civ/\heii]$_{\rm obs}=1.25\pm1.1$.
A Welch $t$-test yields a $p$-value of 28\%, confirming that the two populations are statistically indistinguishable. 
A more accurate assessment of the CGM metallicity requires extended non-resonant emissions, such as H$\alpha$ and \ciii \citep{Guo2020}. 


\section{Conclusions} \label{sec:conclusion}
We have presented a comprehensive analysis of Ly$\alpha$ nebulae around 59 unobscured and 26 obscured quasars at $z=2-3$, observed with Keck/KCWI.
By systematically comparing the morphology, surface brightness profiles, and kinematics of nebulae between the two populations, we have demonstrated how CGM nebulae can be used to constrain the high-$z$ AGN evolutionary picture.
Our main conclusions are as follows:

\begin{itemize}

\item Ly$\alpha$ nebulae surrounding unobscured quasars show notably lower symmetry ($\alpha_{\rm w}=0.2-0.6$) than those around obscured quasars ($\alpha_{\rm w}=0.6-0.8$).
This behavior runs counter to the unified model expectation, which predicts that nebulae viewed face-on through the ionization cone—as is the case for unobscured quasars—should appear more symmetric (Sec.~\ref{nebular_morphology}). 

\item The Ly$\alpha$ nebulae of unobscured quasars are more spatially extended, exhibiting $r_{\rm h}=10.7\pm0.5$ kpc versus $r_{\rm h}\approx6.6-7.7$ kpc for obscured quasars.
This too contradicts the unified-model prediction, which foresees steeper profiles for unobscured quasars ($r_{\rm h,sim}=10.5$ kpc) and shallower ones for obscured quasars ($r_{\rm h,sim}=12.2$ kpc) (Sec.~\ref{nebular_SB}).

\item Unobscured quasars display a pronounced decline in velocity dispersion in the CGM ($-4.3\pm0.4$ km s$^{-1}$ kpc$^{-1}$), whereas obscured quasars show nearly flat profiles ($-0.6$ \& $-0.2$ km s$^{-1}$ kpc$^{-1}$).
This suggests that large-scale outflows have propagated through the CGM of unobscured quasars but have not yet reached comparable scales in the obscured population (Sec.~\ref{nebular_velocity_dispersion}).

\item Nebular extent and symmetry correlate strongly with the degree of quasar obscuration ($r-W_{4}$; $|R_{\rm s}|\approx0.7$, $p\leq2.5\%$), but show no significant dependence on SMBH mass or radio loudness.
This indicates that obscuration is the primary driver behind the observed nebular variations (Sec.~\ref{nebulae_AGN_connection}). 

\item Simulations under the AGN unified model yield only a $3.2\times10^{-3}\%$ probability of simultaneously reproducing the observed differences in both nebular morphology and extent. 
Even when AGN feedback is considered, the predicted nebular contrast between the two populations is likely enhanced rather than diminished, indicating the unified scheme is challenging to reproduce these observations (Sec.~\ref{unified_model_comparison}). 

\item The evolutionary scenario naturally accounts for the above findings: as large-scale AGN feedback drives the transition from an obscured to an unobscured phase, it redistributes gas to larger radii, introduces anisotropy, and drives turbulence, producing nebulae around unobscured quasars that are more extended, more asymmetric, and exhibit strong velocity dispersion gradients.

\item The \civ/Ly$\alpha$ and \civ/\heii \ line ratios measured within a $15-30$ kpc annulus show no significant difference between the two quasar populations, suggesting comparable CGM metallicities.
However, the resonant nature of Ly$\alpha$ and \civ \ limits the reliability of this diagnostic; firm conclusions regarding the metallicity prediction of the evolutionary scenario will require observations of non-resonant emission lines (Sec.~\ref{line_ratio}).

\end{itemize}


Taken together, our results provide strong evidence that the standard orientation-based unified model cannot fully account for the diverse nebular properties observed around quasars at $z=2-3$. 
Instead, the findings favor an evolutionary scenario in which the SMBH and the large-scale CGM are tightly coupled.
The redshift range $z=2-3$ coincides with the peak of cosmic star formation \citep{Madau2014}, sitting between the epoch of abundant LRDs at $z>4$ \citep{Matthee2024,Harikane2023,Maiolino2024} and the well-established unified picture of quasars in the local Universe.
Quasars at this transitional epoch may therefore represent a critical stage in SMBH growth, fundamentally tied to how massive AGN-hosting halos accrete gas from the IGM. 
Characterizing the nature of this transitional epoch could yield essential insights into SMBH growth and its role in shaping galaxies across cosmic time.

\begin{acknowledgments}
SZ is funded by the China National Postdoctoral Program for Innovative Talents. 
AO’s contribution to this project was made possible by funding from the Carl Zeiss Foundation.
\end{acknowledgments}

%

\vspace{5mm}
\facilities{HST(STIS), Swift(XRT and UVOT), AAVSO, CTIO:1.3m,
CTIO:1.5m,CXO}





\appendix

\bibliography{sample631}{}
\bibliographystyle{aasjournal}



\end{document}